\documentclass[review, nopreprintline]{elsarticle}

\usepackage{lineno}
\modulolinenumbers[5]

\usepackage{framed,multirow}
\usepackage{amssymb}
\usepackage{latexsym}
\usepackage{chemformula} 
\usepackage{amsmath}
\usepackage{bm} 
\usepackage{url}

\DeclareMathOperator*{\cdf}{cdf}       

\usepackage[T1]{fontenc}
\usepackage[utf8]{inputenc} 

\begin{document}

\thispagestyle{plain}
\begin{Large}
\noindent \textbf{IEEE Copyright Notice}
\vspace{0.4cm}
\end{Large}

\small{
\noindent Copyright (c) 2023 IEEE

\noindent Personal use of this material is permitted. Permission from IEEE must be obtained for all other uses, in any current or future media, including reprinting/republishing this material for advertising or promotional purposes, creating new collective works, for resale or redistribution to servers or lists, or reuse of any copyrighted component of this work in other works.
}

\hfill

\noindent Published in: IEEE Transactions on Radiation and Plasma Medical Sciences, May 2023

\hfill

\noindent DOI: 10.1109/TRPMS.2023.3243735

\hfill

\hfill

\noindent Cite as:

\hfill

\fbox{
\parbox{40em}{
\footnotesize{
R. Y. Shopa et al., "TOF MLEM Adaptation for the Total-Body J-PET With a Realistic Analytical System Response Matrix," in IEEE Transactions on Radiation and Plasma Medical Sciences, vol. 7, no. 5, pp. 509-520, May 2023, doi: 10.1109/TRPMS.2023.3243735.}
}
}

\hfill

\noindent BibTex:

\hfill

\fbox{
\parbox{40em}{
\footnotesize{
$@$ARTICLE\{10041212,

author=\{Shopa, R.Y. and Baran, J. and Klimaszewski, K. and Krzemie{\'{n}}, W. and Raczy{\'{n}}ski, L. and Wi{\'{s}}licki, W. and Brzezi{\'{n}}ski, K. and Chug, N. and Coussat, A. and Curceanu, C. and Czerwi{\'{n}}ski, E. and Dadgar, M. and Dulski, K. and Gajewski, J. and Gajos, A. and Hiesmayr, B. C. and Valsan, E. Kavya and Korcyl, G. and Kozik, T. and Kumar, D. and Kap{\l}on, {\L}. and Moskal, G. and Nied{\'{z}}wiecki, S. and Panek, D. and Parzych, S. and P{\'{e}}rez del Rio, E. and Ruci{\'{n}}ski, A. and Sharma, S. and Shivani, S. and Silarski, M. and Skurzok, M. and St{\c{e}}pie{\'{n}}, E. and Ardebili, F. Tayefi and Ardebili, K. Tayefi and Moskal, P.\},
 
 journal=\{IEEE Transactions on Radiation and Plasma Medical Sciences\}, 
 
title=\{TOF MLEM Adaptation for the Total-Body J-PET With a Realistic Analytical System Response Matrix\}, 

year=\{2023\},

volume=\{7\},

number=\{5\},

pages=\{509-520\},

doi=\{10.1109/TRPMS.2023.3243735\}\}
}
}
}

\begin{frontmatter}

\title{TOF MLEM Adaptation for the Total-Body J-PET\\ with a Realistic Analytical System Response Matrix}

\author[1]{R.Y.~Shopa\corref{cor1}}
\ead{Roman.Shopa@ncbj.gov.pl}
\cortext[cor1]{Corresponding author at: Department of Complex Systems, National Centre for Nuclear Research, 05-400 Otwock-{\'{S}}wierk, Poland 
  Tel.: +48-22-273-13-06;  
  fax: +48-22-273-16-87;}
\author[2,3,4]{J.~Baran} 
\author[1]{K.~Klimaszewski}
\author[3,4,5]{W. Krzemie{\'{n}}}
\author[1]{L.~Raczy{\'{n}}ski}
\author[1]{W.~Wi{\'{s}}licki}
\author[6]{K.~Brzezi{\'{n}}ski}
\author[2,3,4]{N.~Chug}
\author[2,3,4]{A.~Coussat}
\author[7]{C.~Curceanu}
\author[2,3,4]{E.~Czerwi{\'{n}}ski}
\author[2,3,4]{M.~Dadgar}
\author[2,3,4]{K.~Dulski}
\author[6]{J.~Gajewski}
\author[2,3,4]{A.~Gajos}
\author[8]{B.C.~Hiesmayr}
\author[2,3,4]{E.~Kavya Valsan}
\author[2,3,4]{G.~Korcyl}
\author[2,3,4]{T.~Kozik}
\author[2,3,4]{D.~Kumar}
\author[2,3,4]{{\L}.~Kap{\l}on}
\author[3,4,9]{G.~Moskal}
\author[2,3,4]{S.~Nied{\'{z}}wiecki}
\author[2,3,4]{D.~Panek}
\author[2,3,4]{S.~Parzych}
\author[2,3,4]{E.~P{\'{e}}rez del Rio}
\author[6]{A.~Ruci{\'{n}}ski}
\author[2,3,4]{S.~Sharma}
\author[2,3,4]{Shivani}
\author[2,3,4]{M.~Silarski}
\author[2,3,4]{M.~Skurzok}
\author[2,3,4]{E.~St{\c{e}}pie{\'{n}}}
\author[2,3,4]{F.~Tayefi Ardebili}
\author[2,3,4]{K.~Tayefi Ardebili}
\author[2,3,4]{P.~Moskal} \fnref{fn1}
\fntext[fn1]{P. Moskal is a senior author.}

\address[1]{Department of Complex Systems, National Centre for Nuclear Research, Otwock-{\'{S}}wierk, Poland}
\address[2]{Faculty of Physics, Astronomy and Applied Computer Science, Jagiellonian University, Krak\'{o}w, Poland}
\address[3]{Total-Body Jagiellonian-PET Laboratory, Jagiellonian University, Krak\'{o}w, Poland}
\address[4]{Center for Theranostics, Jagiellonian University, Krak\'{o}w, Poland}
\address[5]{High Energy Physics Division, National Centre for Nuclear Research, Otwock-{\'{S}}wierk, Poland}
\address[6]{Institute of Nuclear Physics, Polish Academy of Sciences, Krak\'{o}w, Poland}
\address[7]{INFN, Laboratori Nazionali di Frascati, Frascati, Italy}
\address[8]{Faculty of Physics, University of Vienna, Vienna, Austria}
\address[9]{Faculty of Chemistry of the Jagiellonian University, Krak\'{o}w, Poland}

\begin{abstract}
We report a study of the original image reconstruction algorithm based on the time-of-flight maximum likelihood expectation maximisation (TOF MLEM), developed for the total-body (TB) Jagiellonian PET (J-PET) scanners. The method is applicable to generic cylindrical or modular multi-layer layouts and is extendable to multi-photon imaging. The system response matrix (SRM) is represented as a set of analytical functions, uniquely defined for each pair of plastic scintillator strips used for the detection. A realistic resolution model (RM) in detector space is derived from fitting the Monte Carlo simulated emissions and detections of annihilation photons on oblique transverse planes. Additional kernels embedded in SRM account for TOF, parallax effect and axial smearing. The algorithm was tested on datasets, simulated in GATE for the NEMA IEC and static XCAT phantoms inside a 24-module 2-layer TB J-PET. Compared to the reference TOF MLEM with none or a shift-invariant RM, an improvement was observed, as evaluated by the analysis of image quality, difference images and ground truth metrics. We also reconstructed the data with additive contributions, pre-filtered geometrically and with non-TOF scatter correction applied. Despite some deterioration, the obtained results still capitalise on the realistic RM with better edge preservation and superior ground truth metrics. The envisioned prospects of the TOF MLEM with analytical SRM include its application in multi-photon imaging and further upgrade to account for the non-collinearity, positron range and other factors.
\end{abstract}


\end{frontmatter}
This work did not involve human subjects or animals in its research.

\section{Introduction}
As an established diagnostic tool in nuclear medicine, positron emission tomography (PET) remains a rapidly evolving technology \cite{Alavi2022} in regards to general concepts \cite{Moskal2021, Gajos2021Nature, Cherry2018, Tsuda2008, Marcinkowski2016, Karp2020a, Karakatsanis2022}, new detector materials \cite{Gundacker2013, Moskal2014, Kaplon2021, Lecoq2022, Pagano2022, Mohr2022}, radioactive tracers \cite{Carter2020, Matulewicz2022, Choinski2022}, readout methods \cite{Marcinkowski2016, Palka2017, Korcyl2018, Yoshida2020, Kwon2021} or hybrid imaging \cite{Mehranian2017, Ote2022}. One of the primary trends is the development of total-body (TB) PET systems with the axial field-of-view (AFOV) reaching up to $2$ metres, which superior sensitivity and time-of-flight (TOF) information available make them suitable for new diagnostic methods \cite{Badawi2019}. Modern clinical TB scanners achieve the coincidence resolving time (CRT) of $210$~ps and spatial resolution $\sim3$ mm \cite{Karp2020a, VanSluis2019, Spencer2021, Alberts2021, Prenosil2022}, and even better numbers are reported for small animal PET tomographs or detectors that account for the depth of interaction \cite{Tsuda2008, Marcinkowski2016, Yoshida2020}. 
Such advancements constitute demanding requirements for tomographic image reconstruction \cite{Efthimiou2020}. Besides, TOF TB produces terabytes of data during acquisition and operates with expanded projection (bin-) space due to a much higher number of detection elements \cite{Zhou2014, Raczynski2020}. Novel deep learning (DL) algorithms provide remarkable achievements \cite{Haggstrom2019, Whiteley2021, Mehranian2021, Ote2022}, but also consume colossal computational resources \cite{Reader2021}. At the same time, classical iterative methods -- such as maximum likelihood expectation maximisation (MLEM) \cite{Shepp1982} -- not only reflect the PET physics model for radio-tracers, but remain a reliable solution for regular machinery \cite{Reader2021a}. Moreover, while MLEM, as the simplest iterative method, can serve as a good benchmark for testing new tomographs, many DL techniques are successfully embedded in such algorithms \cite{Mehranian2021, Gong2019, Reader2021a}.

To define a fair system response matrix (SRM) -- a key element of iterative algorithms -- for a modern TB PET scann\-er, one must take into account multiple factors: detector blur, non-collinearity, parallax effect or positron range, especially in multi-photon acquisition mode, where isotopes with wider kinetic energy spectra are used \cite{LeLoirec2007, Moses2011, Moskal2019, Carter2020}. Another technical challenge for multi-photon PET is similar to TB -- a massive yet sparse bin space, which renders the sinogram data format inefficient and makes it hard to assess calibration, sensitivity and attenuation factors. That favours the so-called list-mode MLEM, based on the event-by-event data processing and easy to scale for multi-threading \cite{Barrett1997}. 

The SRM definition is a fairly difficult task for the Jagiellonian PET (J-PET) -- an innovative cost-effective technology which utilises Compton scattering for the detection of electron-positron ($e^-e^+$) annihilation photons and can operate in multi-photon mode, as well \cite{Moskal2014, Raczynski2014, Moskal2016, Palka2017, Korcyl2018, Moskal2020, Moskal2021, Gajos2021Nature}. One J-PET element is composed of an axially aligned plastic scintillator strip and two photomultipliers (PMs) attached at its ends. The PMs collect optical photons (a result of a scatter that happened inside the strip) as time signals, giving hit position and TOF. A combination of silicon PMs (SiPM) and EJ-230 scintillators could achieve $\text{CRT}=237$~ps for a $2$-m long strip \cite{Moskal2021a}. A multi-layer modular arrangement is beneficial to compensate for the lower sensitivity and worse axial resolution, improved by attaching an additional array of wavelength shifting (WLS) scintillators \cite{Smyrski2017}. Arguably, such a complex geometry would produce a detector blur and additional scatter that affect SRM. Early application of a generic TOF MLEM from an open-source CASToR software resulted in noisy images and lower resolution, inferior to the algorithms designed specifically for the J-PET \cite{Raczynski2020, ShopaRAP2021}.

The main goal of this work is to conduct a comprehensive simulation study of the recently developed TOF MLEM that defines SRM analytically for the multi-layer TB J-PET \cite{ShopaRAP2021}. We upgrade the SRM model to account for the parallax effect and tested it using the data for the NEMA IEC and the eXtended CArdiac-Torso (XCAT) phantoms, simulated inside a modular TB J-PET using the GATE framework \cite{Jan2004}. The research also reports the first results for the J-PET with scatter correction applied.

\section{Methodology}

\subsection{System response matrix and its decomposition}
The relation between the expectation $\langle\textbf{y}\rangle$ of the PET data and the unknown radioactive tracer distribution $\bm\lambda$ is defined by a linear model:
\begin{equation}
\label{Eq:ForwardProjGenVector}
\langle\textbf{y}\rangle = \textbf{M}\bm\lambda + \hat{\textbf{b}},
\end{equation}  

\noindent where $\textbf{M}$ denotes the SRM of a PET system and $\hat{\textbf{b}}$ is the observation error, constituted by random and scatter coincidences \cite{Lange1984}. $\textbf{M}$ is sparse and can be decomposed in order of physical phenomena:
\begin{equation}
\label{Eq:SRMDecompose}
\textbf{M}=\textbf{M}_\text{sens}\textbf{M}_\text{det.res}\textbf{M}_\text{att}\textbf{M}_\text{geom}\textbf{M}_\text{img.res},
\end{equation}  

\noindent where $\textbf{M}_\text{sens}$ represents the sensitivity, $\textbf{M}_\text{det.res}$ -- blur imposed in detector space, $\textbf{M}_\text{att}$ -- attenuation factors, $\textbf{M}_\text{geom}$ -- geometric projection matrix and $\textbf{M}_\text{img.res}$ -- image space resolution, such as positron range. The elements of $ m_{ij}\in\textbf{M}$ reflect the conditional probabilities for detecting the coincident emissions from some $j$-th voxel by $i$-th bin (pair of detectors for the case of $511$-keV back-to-back annihilation photons).

\subsection{List-mode TOF MLEM}
Assuming the mean of the measurements is an unbiased estimate, 
the data can be modelled by the Poisson distribution \cite{Shepp1982}. A log-likelihood cost function is utilised considering the discrete photon counts $\hat{\textbf{y}}\equiv\lbrace\hat{y}_i\rbrace$ and the unknown radio-tracer distribution $\bm{\lambda} \equiv \lbrace \lambda_j \rbrace$. The reconstructed image $\langle\bm{\lambda}\rangle$ is acquired via maximisation:
\begin{equation}
\label{Eq:LogLikelihoodMaximisation}
\langle\bm{\lambda}\rangle = \arg\max\limits_{\bm{\lambda}}L(\bm{\lambda}, \hat{\textbf{y}}),
\end{equation}   
\begin{equation}
\label{Eq:LogLikelihood}
L(\bm{\lambda}, \hat{\textbf{y}}) = \sum\limits_{i \in \mathcal{I}} (-\langle y_i\rangle + \hat{y}_i \log{\langle y_i\rangle}),
\end{equation}   

\noindent where $\langle y_i\rangle\in\langle \textbf{y}\rangle$ is the expectation value for the $i$-th data bin. An iterative MLEM solution to (\ref{Eq:LogLikelihoodMaximisation}) is given by \cite{Shepp1982, Lange1984}:
\begin{equation}
\label{Eq:MLEMini}
\lambda^{(k+1)}_j = \frac{\lambda^{(k)}_j}{\sum\limits_{i\in\mathcal{I}}m_{ij}}
\sum\limits_{i\in\mathcal{I}}
{
	\frac{m_{ij} \hat{y}_i}
	{\sum\limits_{j'\in\mathcal{J}} m_{ij'}\lambda_{j'}^{(k)} + \hat{b}_i }
},
\end{equation}    

\noindent where $k$ is the iteration number, $\hat{b}_i = \hat{r}_i + \hat{s}_i$ are the elements of the error $\hat{\textbf{b}}$ ($\hat{r}_i$ and $\hat{s}_i$ -- the contributions of random and scatter coincidences, respectively). 

In practice, the detector sensitivity $\textbf{M}_\text{sens}$ depends on both geometric and calibration factors, but we shall take into account only the former for the simulated data. For convenience, the following notation for the SRM elements will be used:
\begin{equation}
\label{Eq:SRMThreeComponent}
m_{ij} = n_i a_i \chi_{ij},
\end{equation}

\noindent where $n_i\in\textbf{M}_\text{sens}$, $a_i\in\textbf{M}_\text{att}$ and $\chi_{ij}$ is the shift-variant SRM part, approximating the original model (\ref{Eq:SRMDecompose}) by combining $\textbf{M}_\text{det.res}$ and $\textbf{M}_\text{geom}$, while moving the attenuation operation $\textbf{M}_\text{att}$ to the left to be applied after the detector resolution model. $\textbf{M}_\text{det.res}$ and $\textbf{M}_\text{geom}$ were similarly combined in \cite{Matej2009}, only the attenuation operation was moved more accurately to the right to be performed before the SRM operation, as allowed by the histo-image data partitioning framework.

Following common practice for TOF PET scanners, we use the list-mode MLEM, where the main sum over bins in (\ref{Eq:MLEMini}) is replaced by a sum over each measured coincident event $\epsilon\in\mathcal{E}$ \cite{Barrett1997}. Considering (\ref{Eq:SRMThreeComponent}), we could rewrite (\ref{Eq:MLEMini}) as follows:
\begin{equation}
\label{Eq:TOFMLEMids}
\lambda^{(k+1)}_j = \frac{\lambda^{(k)}_j}{\sum\limits_{i\in\mathcal{I}}{n_i a_i \chi_{ij}} }
\sum\limits_{\epsilon\in\mathcal{E}}
{
	\frac{n_{i_\epsilon} \chi_{(it)_\epsilon, j}}
	{\sum\limits_{j'\in\mathcal{J}_\epsilon} n_{i_\epsilon} \chi_{(it)_\epsilon, j'}\lambda_{j'}^{(k)} +  \hat{b}_{i_\epsilon}^{*}}
}.
\end{equation} 

\noindent where $\hat{b}_{i_\epsilon}^{*} = (\hat{r}_{i_\epsilon} + \hat{s}_{i_\epsilon})/ a_{i_\epsilon}$. The TOF information is embedded in SRM using the tensor form of the time-spread function $K_{itj}$ as $\chi_{it,j} = \chi_{ij}K_{itj}$ (under condition $\sum_{t\in\mathcal{T}_i} \chi_{it,j}=\chi_{ij}$) \cite{Efthimiou2019}.

\subsection{Analytical SRM for J-PET}
\begin{figure}[!t]
\centering
\includegraphics[width=22pc]{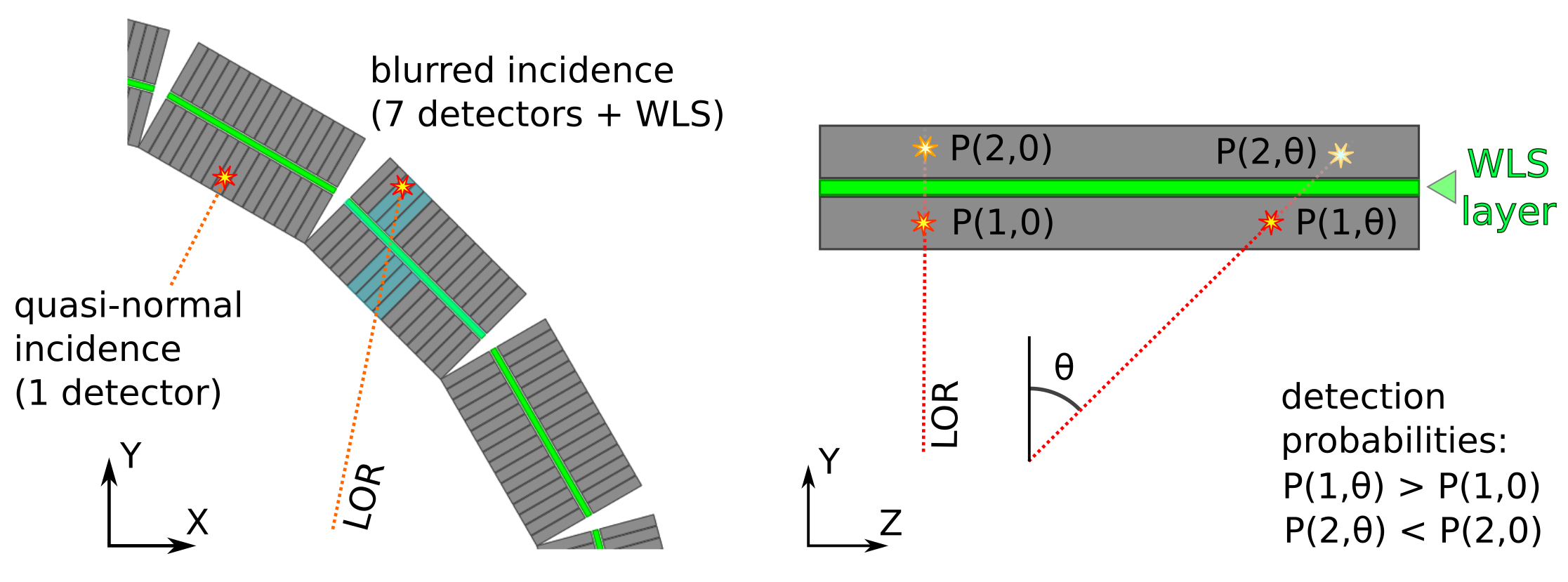}
\caption{Hits in TB J-PET: scintillators are in grey and WLS -- in green \cite{Moskal2021a}.}
\label{Fig:DJPET}
\end{figure}

The detector RM $\textbf{M}_\text{det.res}$ is difficult to estimate for multi-layer modular TB J-PET scanners, since a photon may cross many scintillators before being scattered (Fig.~\ref{Fig:DJPET}, left). On the other hand, a unique feature is a continuous character of the J-PET strips in the axial direction, so we can assume the detection probabilities depend smoothly on the obliqueness angle $\theta$ (Fig.~\ref{Fig:DJPET}, right). 

Extending the idea from \cite{Strzelecki2016}, a number of 2D simulations were conducted on the transverse plane for back-to-back emissions and detections in a J-PET scanner. Each such simulation was launched for a different $\theta$, increased by a  step $\pi/16$ starting from $0$, while the attenuation coefficient $\mu=0.096\,\text{cm}^{-1}$ (for $511$-keV photons in EJ-230 scintillator \cite{Moskal2021a}) was adjusted accordingly as $\mu/\cos\theta$. As a result, the counts for each $\theta$ reflected not only the detector blur, but also the geometrical contribution to the sensitivity factors $n_i$. 

\begin{figure}[!t]
\centering
\includegraphics[width=21pc]{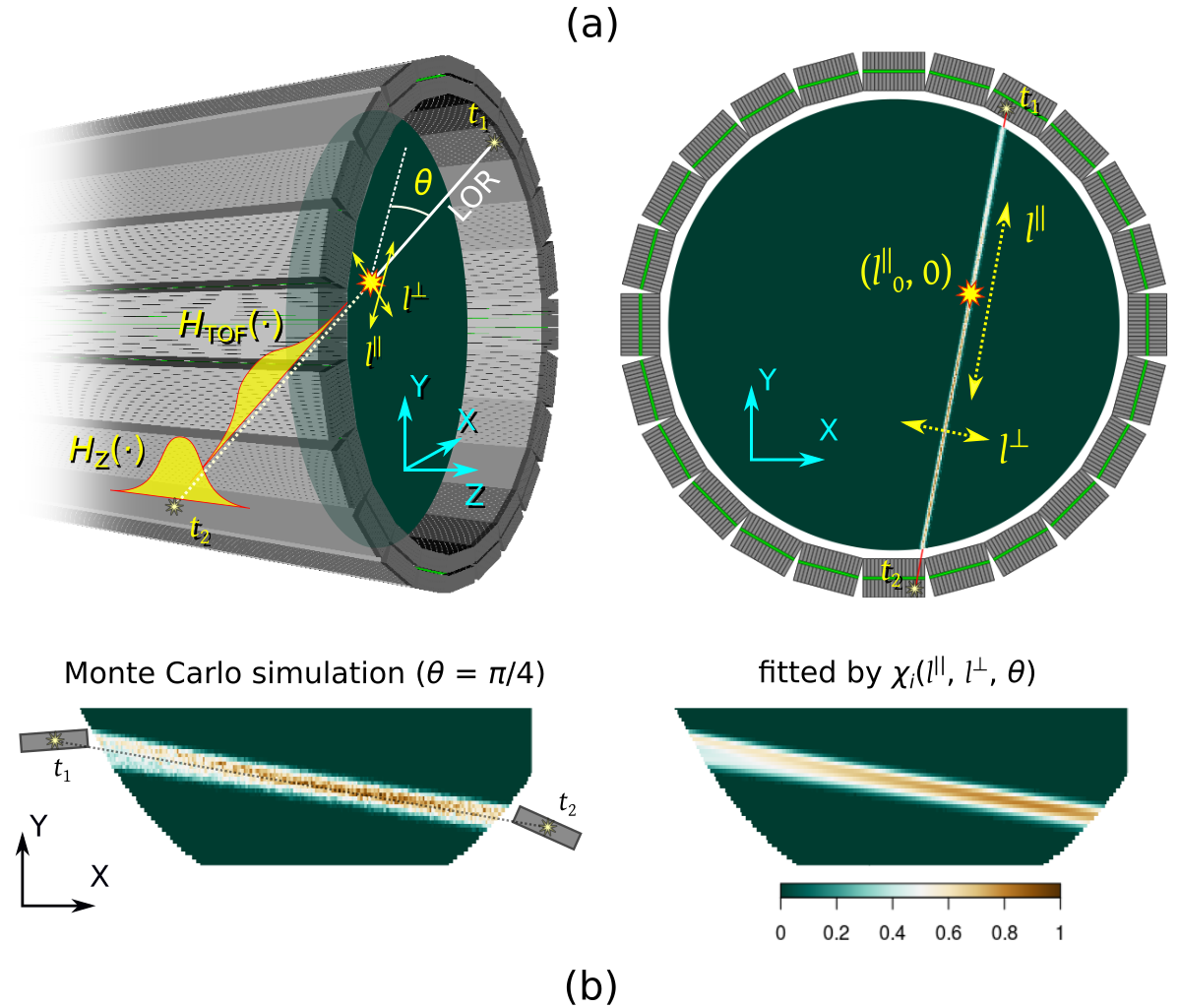}
\caption{Schematic depiction of the variables and kernels used in the analytical RM of the modular J-PET (a); simulated and fitted detection probability distribution for an exemplary bin (b, strip IDs 474 and 666). The point $(l^\parallel=l^\parallel_0, l^\perp=0)$ corresponds to the point of the annihilation.} 
\label{Fig:SRMinDJPET}
\end{figure}

At the following stage, the simulated data were split into pairs of J-PET strips, which reduced the bin space to $\mathcal{I}\subset\mathbb{R}^2$. Next, a phenomenological fitting function $\chi_i\left(\cdot\right)$ was applied for each bin, to substitute the joint matrix $\chi_{ij}$ and $n_i$. To complete the probabilistic model, we used two additional kernels: $H_\text{TOF}(\cdot)$ -- to account for the TOF information -- and $H_Z(\cdot)$ -- for hit positions (Compton scatterings in scintillators) being smeared along $Z$-axis (Fig.~\ref{Fig:SRMinDJPET}, a). 

It is not practical to use the Cartesian voxel coordinates and times of hits ($t_1$ and $t_2$ in Fig.~\ref{Fig:SRMinDJPET}) as the arguments of the aforementioned functions. We therefore defined new variables with respect to a line-of-response (LOR) -- the transverse coordinates $l^\parallel$ and $l^\perp$ along LOR and in the normal direction, respectively. That allowed to employ a log-polynomial of the $5$-th order as the fitting functions $\chi_i\left(l^\parallel, l^\perp, \theta \right)$ (Fig.~\ref{Fig:SRMinDJPET}, b). The full expression is given in the Appendix.
TOF and $Z$-kernels can be expressed using the Gaussian cumulative distribution function (CDF) -- $\cdf(x, \mu, \sigma)$, where $\mu$ is mean, $\sigma$ -- standard deviation (SD):
\begin{equation}
\label{Eq:HkernelsCDF}
\begin{split}
H_\text{TOF}(l^\parallel, l^\parallel_0) &= \cdf\left[ l^\parallel_{p+1} - l^\parallel,l^\parallel_0,\sigma_{\text{TOF}}^\parallel \right] - \cdf\left[ l^\parallel_p - l^\parallel,l^\parallel_0,\sigma^\parallel_{\text{TOF}} \right],\\
H_Z(z,l^\parallel)  &= \cdf\left[ z_{q+1}-z,z(l^\parallel),\sigma_Z \right] - \cdf\left[ z_q-z,z(l^\parallel),\sigma_Z \right],
\end{split}
\end{equation} 

\noindent where $l^\parallel$ is calculated from $t_1$ and $t_2$ and lies between the transverse projections $l^\parallel_{p+1}$ and $l^\parallel_p$ of the TOF bins allocated along LOR (similarly to the STIR framework -- see \cite{Efthimiou2019}), $l^\parallel_0$ corresponds to the estimated position of the $e^-e^+$ annihilation and SD $\sigma^\parallel_{\text{TOF}}=\sigma_{\text{TOF}}\cos{\theta}$ depends on CRT \cite{Moskal2016}. Similarly, $z$ is between the voxel coordinates $z_{q+1}$ and $z_q$, while the axial position $z(l^\parallel)$ denotes a point at LOR that corresponds to $l^\parallel$. Note that $H_\text{TOF}(l^\parallel, l^\parallel_0)$ is a transverse projection of the initial oblique TOF-kernel shown in Fig.~\ref{Fig:SRMinDJPET}, a.

TB scanners with large AFOVs require yet another axial correction to account for the parallax effect \cite{Zhang2017}. The additional kernel $H^\text{prlx}_Z(z,l^\parallel,i,\theta)$ depends both on a bin and a voxel and is triangular for the ideal scanner. $H_Z(\cdot)$ is replaced by a convolution $H^\text{CRT}_Z(z^\parallel,l) * H^\text{prlx}_Z(z^\parallel,l,i,\theta)$, where CRT denotes the Gaussian part. 

Eventually, the probabilities $\chi_{(it)_\epsilon, j}$ in (\ref{Eq:TOFMLEMids}) for some event $\epsilon$ are replaced as follows:
\begin{equation}
\label{Eq:SRManalyticalNew}
\begin{split}
\chi_{(it)_\epsilon, j} \rightarrow
	\chi_{i_\epsilon}\left(l^\parallel_j, l^\perp_j, \theta_\epsilon \right) 
	\cdot H_\text{TOF}\left[ l^\parallel_j, (l^\parallel_0)_\epsilon \right]  \cdot
	\left[ H^\text{CRT}_Z(z_j,l^\parallel_j) * H^\text{prlx}_Z(z_j,l^\parallel_j,i_\epsilon,\theta_\epsilon) \right].
\end{split}
\end{equation} 

Since SRM is represented by the fitting parameters, its size is relatively small, and the symmetry of the scanner allows it to be reduced at least eight times -- from $\sim0.4$~Gb to $\sim50$~Mb.

\subsection{GATE simulation setup}
The simulation geometry of the $24$-module TB J-PET replicated our previous study \cite{Moskal2021a} -- to resemble $1.5$-m $6$-ring PennPET Explorer scanner (for NEMA phantom simulations) and $2$-m uExplorer (for the XCAT) \cite{Badawi2019, Karp2020a}. The detector arrangement in one module (see Fig.~\ref{Fig:DJPET}) was as follows: $32$ EJ-230 plastic strips of rectangular $30$~mm $\times$ $6$~mm cross-section, arranged in $2$ layers of radii $408.1$~mm and $443.1$~mm, with a $3$-mm thick mid-layer of WLS scintillators put in between. We show EJ-230 in grey and WLS in green in all figures. 

\begin{figure}[htb]
\centering
\includegraphics[width=21pc]{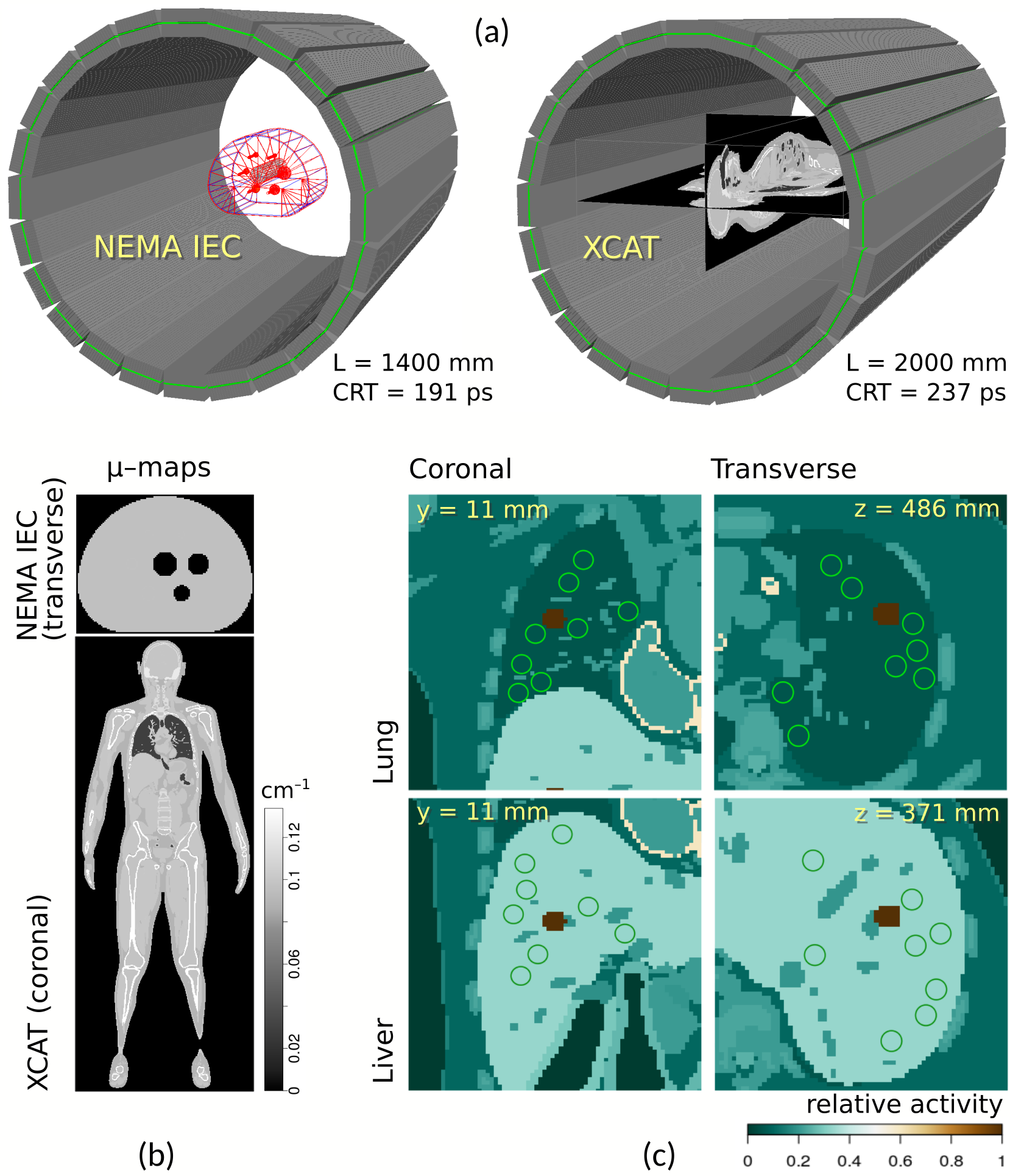}
\caption{Simulated phantoms inside the TB J-PET scanners (a): NEMA IEC (shown as wireframe) and XCAT (as $\mu$-map), cross-sections of the attenuation maps (b), locations of the $12$-mm spherical lesions (solid red circles) and the allocated background ROIs (green circles) in the simulated XCAT (c).} 
\label{Fig:PhantomsMusROIs}
\end{figure}

Two simulations were conducted in GATE -- for a NEMA IEC and a static XCAT phantoms filled with the radioactive water with the dissolved ${}^{18}\text{F-FDG}$ and placed at the centre of the TB J-PET scanners (Fig.~\ref{Fig:PhantomsMusROIs}, a). A dedicated software (see \cite{Segars2018}) was utilised to define the XCAT activity and attenuation ($\mu$-) maps. The simulation parameters are given in Table~\ref{Tab:SimParams}. Total activity and acquisition times for the NEMA IEC were selected to be consistent with our previous works, namely to match either $10^7$ ($35$-s scan -- see \cite{Raczynski2020, Shopa2021, ShopaRAP2021}) or 
$153\times 10^6$ ($500$-s, same dataset as in \cite{Moskal2021a}) true events. As for the XCAT, the value $115$~MBq reflected similar studies made for $25$-MBq phantom in uExplorer \cite{Badawi2019, Zhang2017}, assuming its sensitivity is about $4.6\times$ higher than for the $2$-m TB J-PET (see \cite{Moskal2021a}). The scan duration of $120$~s is in between those for the smaller phantom, considering also longer diagnostics would require motion correction. Peak activity concentrations were about $22$~kBq/ml (NEMA IEC) and $12.5$~kBq/ml (XCAT) -- close to the maximums of the noise equivalent count rate (NECR), estimated in the earlier work for uniform elongated sources of diverse lengths \cite{Moskal2021a}.

\begin{table}[!t]
\renewcommand{\arraystretch}{1.15}
\caption{GATE simulation parameters of the two phantoms\label{Tab:SimParams}}
\centering
\begin{tabular}{lcccc}
\hline 
\hline \\[-5.0ex]
\vspace{0.1cm}	
Phantom	&Scanner length & CRT					& Total activity		& Scan time\\
\hline \\[-5.0ex]
\vspace{0.1cm}	
NEMA IEC 	& $L=1400$~mm 	& $191$~ps 	& $59$~MBq 		& $35$~s, $500$~s\\
\vspace{0.1cm}	
XCAT			& $L=2000$~mm 	& $237$~ps 	& $115$~MBq		& $120$~s\\
\hline \\[-5.0ex]
\end{tabular}
\end{table}

\subsection{Data pre-processing, sensitivity and attenuation correction}
Data pre-selection was performed using the criteria established in previous works: a $3$-ns coincidence window and a $200$-keV threshold for the minimal energy deposited via Compton scattering \cite{Raczynski2020, Moskal2021a}. The majority of the research was done for the data constituted only by true coincidence events. The other studies, which considered the additive contributions $\hat{b}_i>0$,  had some geometrical restrictions imposed on LORs (must cross the phantom) and time differences between hits ($c_0\vert t_1 - t_2 \vert < \text{LOR length}$, $c_0$ -- the speed of light in the air). The distributions of coincidences by type before and after such a procedure are shown in Table~\ref{Tab:CTypes} (the $500$-s scan of the NEMA IEC exhibited similar fractions to that for $35$-s). As explored, the geometrical filtering -- similar to the one used in \cite{Moskal2021} -- does not corrupt the Poisson distribution for the true and scattered events, yet cuts out a large fraction of randoms, especially for the NEMA IEC.

\begin{table}[!t]
\renewcommand{\arraystretch}{1.05}
\caption{GATE coincidence statistics ($\times 10^6$ events)\label{Tab:CTypes}}
\centering
\begin{tabular}{lcccc}
\hline 
\hline \\[-5.0ex]
\multirow{2}{*}{Phantom} & \multirow{2}{*}{True} & Phantom & Detector & \multirow{2}{*}{Random}\\
\vspace{0.1cm}
&&scattered&scattered&\\
\hline \\[-5.0ex]
\multicolumn{5}{c}{NEMA IEC ($35$-s scan)}\\
\multirow{2}{*}{Initial} 	& $10.24$ & $14.67$ & $4.69$ & $14.51$ \\
\vspace{0.1cm}	
&	($23.2\%$) & ($33.2\%$) & ($10.6\%$) & ($32.9\%$) \\
\multirow{2}{*}{Filtered} & $10.11$ & $6.72$ & $4.29$ & $0.55$ \\ 
\vspace{0.2cm}	
&	($46.7\%$) & ($31.0\%$) & ($19.8\%$) & ($2.5\%$) \\
\multicolumn{5}{c}{XCAT}\\
\multirow{2}{*}{Initial} & $42.34$ & $66.72$ & $12.04$ & $141.14$ \\
\vspace{0.1cm}	
&	($16.1\%$) & ($25.4\%$) & ($4.6\%$) & ($53.9\%$) \\
\multirow{2}{*}{Filtered} & $41.03$ & $50.77$ & $10.98$ & $20.66$ \\
\vspace{0.2cm}
&	($33.2\%$) & ($41.1\%$) & ($8.9\%$) & ($16.7\%$) \\
\hline \\[-5.0ex]
\end{tabular}
\end{table}

For the list-mode format, we favoured the performance of the algorithm over data reduction, having stored the scintillator IDs, the expected annihilation coordinates, the obliqueness angle $\theta$ and time differences for two hits rescaled to the variable $l^\parallel$. That amounted to about $40$ bytes per event, or up to $6.5$~Gb for the largest dataset ($153\times 10^6$ records). On resampling, the size can be reduced at least twice. 

Hit positions in J-PET strips were post-smeared along $Z$-axis, using normal distribution with SD $\sigma_Z=2.12$~mm that reflects WLS resolution (full width at half maximum) $\text{FWHM}=5$~mm \cite{Smyrski2017}. Similarly, times of hits underwent Gaussian smearing, depending on CRT, which degrades for longer AFOV \cite{Moskal2021a} (see Table~\ref{Tab:SimParams}).

The sensitivity and attenuation correction (AC) factors were merged in the denominator sum (\ref{Eq:TOFMLEMids}), using the analytical SRM model (\ref{Eq:SRManalyticalNew}), integrated over the obliqueness angle $\theta$:
\begin{equation}
\label{Eq:SensAttSum}
\sum\limits_{i\in\mathcal{I}}{n_i a_i \chi_{ij}} \rightarrow 
\sum\limits_{i\in\mathcal{I}}\int_{\theta}{a_{i,\theta} \cdot \chi_i\left(l^\parallel_j, l^\perp_j, \theta \right) d\theta}.
\end{equation}  

The factors $a_{i,\theta}$ were calculated by accelerated Siddon projectors \cite{Jacobs1998} using the predefined $\mu$-maps (Fig.~\ref{Fig:PhantomsMusROIs}, b). We imply that the functions $\chi_i(\cdot)$ on the right side of (\ref{Eq:SensAttSum}) have the geometrical contributions $n_i$ already embedded within.

\subsection{Scatter correction}
To better analyse the impact of the RM model, we mainly focused on the subsets with true coincidences, so that $b_{i_\epsilon}^*=0$. However, several reference tests were also launched with the additive factors included (see Table~\ref{Tab:CTypes}). There is an ongoing study dedicated to the advanced classification of the random events in J-PET, hence we concentrated on the scatter correction inside the phantoms. On utilising the scripts for the single scatter simulation (SSS) from the STIR reconstruction framework\footnote[1]{STIR version 3.0 was used.} \cite{Polycarpou2011, Thielemans2012}, low-resolution SSS-sinograms ($48$ views, axial ring difference $\Delta z = 66$~mm) were generated for each phantom, considering the ideal cylindrical scanner geometry that matched the inner radius of the modular J-PET. The list-mode scatter contributions $\hat{s}_{i_\epsilon}$ were estimated using interpolation and divided by the AC factors $a_{i_\epsilon}$. Since STIR does not account for bin sensitivities (required to properly assess additive corrections), the values were also normalised with respect to the SRM model (\ref{Eq:SRManalyticalNew}), i.e. the sum $\sum_{j\in\mathcal{J}_\epsilon} \chi_{i_\epsilon}(\cdot)H_\text{TOF}(\cdot)H_Z(\cdot)$. 

\subsection{Reference TOF MLEM}
The reconstructed activity concentrations were compared with the results, obtained by a list-mode TOF MLEM from the CASToR framework \cite{Merlin2018}, which is easier to utilise for the multi-layer total-body J-PET than STIR. It computes SRM using line projectors with no default model for $\textbf{M}_\text{det.res}$. For consistency with AC, we have selected the accelerated Siddon projector. CASToR has no internal features to calculate detector normalisation factors and scatter correction. The first issue was resolved by providing Monte Carlo simulated sensitivity maps (for two scanner lengths), later converted to the list-mode as required by the framework. As there is no SSS implementation in CASToR, only the data for true events was reconstructed and compared with our results.

\subsection{Image quality and ground truth metrics}
According to NEMA, image quality parameters -- contrast recovery coefficient (CRC) and background variability (BV) -- describe qualitatively the clinical imaging conditions, using the pre-defined circular regions-of-interest (ROIs), which encircle the six NEMA IEC spheres on the transverse plane, intersecting their centres \cite{NEMA2012}. On the background area, twelve additional ROIs are drawn concentrically for each sphere, replicated at $\pm10$~mm and $\pm20$~mm from the main plane. 

The coefficients $\text{CRC}_{H,d}$ and $\text{CRC}_{C,d}$ for a hot (higher activity) or a cold (no activity) sphere of a diameter $d$ are estimated as follows:
\begin{equation}
\text{CRC}_{H,d} = \frac{\frac{\mu_{H,d}}{\mu_{B,d}}-1}{\alpha-1},\quad 
\text{CRC}_{C,d} = 1-\frac{\mu_{C,d}}{\mu_{B,d}},
\label{eq:CRCs}
\end{equation}

\noindent where $\mu_{H,d}$, $\mu_{C,d}$ and $\mu_{B,d}$ are the average intensities of the corresponding ROI around a hot or a cold sphere and on the background, respectively; $\alpha$ is the activity ratio between the hot regions and the background (local contrast).

BV of a sphere depends on the SD of the background $\sigma_{B,d}$:
\begin{equation}
\text{BV}_{d} = \frac{\sigma_{B,d}}{\mu_{B,d}},
\label{eq:BVs}
\end{equation}
\begin{equation}
\sigma_{B,d} = \sqrt{\frac{1}{N_{\text{ROI}}-1}\sum_{i=1}^{N_{\text{ROI}}} (\mu_{B,d,i}-\mu_{B,d})^2},\, N_{\text{ROI}}=60.
\label{eq:SDROI}
\end{equation}

For our simulation, four smaller spheres of the NEMA IEC (diameters $10$~mm, $13$~mm, $17$~mm, $22$~mm) were set hot with $4\times$ higher activity than in the background, while the other two ($28$~mm, $37$~mm) were cold. To emulate the NEMA requirement of averaging (\ref{eq:CRCs})-(\ref{eq:BVs}) over multiple measurements, we randomly drew $30$ sets of background ROIs. 

In a similar way, two spherical $12$-mm lesions were simulated in lung and liver regions of the XCAT, located at $(x,y,z)=$ ($90$~mm, $11$~mm, $486$~mm) and $(x,y,z)=$ ($90$~mm, $11$~mm, $371$~mm), respectively (Fig.~\ref{Fig:PhantomsMusROIs}, c). Local contrasts were set as $16:1$ (lung) and $3:1$ (liver), similarly to the work \cite{Zhang2017}. To estimate CRC and BV, ROIs were drawn around the lesions and on a nearby background. Due to a lack of area with uniform activity, their number was reduced to $8$ (or $40$ in total). To further expand the study, image quality analysis was conducted using two sets of ROIs, drawn on transverse ($XY$) and coronal ($XZ$) planes (Fig.~\ref{Fig:PhantomsMusROIs}, c).

As additional metrics, we used the root mean square error (RMSE) and structural similarity (SSIM) index between the reconstructed image $\langle\bm{\lambda}\rangle$ and the initial radio-trace distribution $\bm{\lambda}^\text{(GT)}$ (ground truth -- GT) defined for the simulation \cite{Wang2004}:
\begin{equation}
\text{RMSE}_\text{GT}=\sqrt{\frac{1}{N_\text{vox}}\sum\limits_{j=1}^{N_\text{vox}}\left( \langle\lambda_j\rangle - \lambda_j^\text{(GT)} \right)^2},
\label{Eq:RMSE}
\end{equation}
\begin{equation}
\label{Eq:SSIM}
\text{SSIM}_\text{GT} = 
\frac{ \left( 2\mu_\lambda\mu_\text{GT}+c_1 \right)
       \left( 2\sigma_\text{cov}+c_2 \right) }
     { \left( \mu_\lambda^2+\mu_\text{GT}^2+c_1 \right) \left(\sigma^2_\lambda+\sigma^2_\text{GT}+c_2 \right)},
\end{equation}   

\noindent where $N_\text{vox}$ is the total number of voxels, $c_1=(0.01L)^2$, $c_2=(0.03L)^2$ -- the		 constants that depend on the peak intensity $L$ of the two images, $\mu_\lambda$, $\mu_\text{GT}$ are the averages, $\sigma_\lambda^2$, $\sigma_\text{GT}^2$ -- the variances, $\sigma_\text{cov}$ -- the covariance between $\langle\bm{\lambda}\rangle$ and $\bm{\lambda}^\text{(GT)}$, respectively. The ideal RMSE is zero, SSIM index -- one.

\subsection{Programming tools and computational resources}
For simulations and analysis, we used {\small\textsf{R/Rcpp}} libraries with vectorized functions \cite{RCoreTeam2020, Eddelbuettel2011}. The executables were launched in the multi-thread mode on $48$ regular nodes of CI{\'{S}} cluster\footnote[2]{CI{\'{S}} -- Centrum Informatyczne {\'{S}}wierk: \url{https://www.cis.gov.pl/}} (Intel Xeon E5-2680v2, 128~GB RAM).

\section{Results}
We modelled SRM and reconstructed the data for the $2.5$~mm $\times$ $2.5$~mm $\times$ $2.5$~mm voxel. The SDs for the kernels (\ref{Eq:HkernelsCDF}) reflected the CRT and axial resolution of WLS: $\sigma_\text{TOF}=c_0\cdot\text{CRT}/(4\sqrt{2\log{2}})$ ($12.2$~mm and $15.1$~mm for the shorter and longer J-PET, respectively), $\sigma_Z=2.12$~mm. TOF bin difference in (\ref{Eq:HkernelsCDF}) was set below $\sigma_\text{TOF}$ at $l^\parallel_{p+1}-l^\parallel_p=5.0$~mm.	

Each reconstructed image was rescaled to kBq/ml: normalised so that its sum matched the true activity of the phantom (Table~\ref{Tab:SimParams}) and later divided by the volume of the voxel.

We shall use the following notation for TOF MLEM:
\begin{itemize}
\item[--] "no RM": the reference algorithm from CASToR which does not consider the detector and axial smearing;
\item[--] "image-based PSF": the CASToR implementation with an image domain shift-invariant RM kernel \cite{Zhou2014, Merlin2018}. A point spread function (PSF) is modelled as a Gaussian convolutional filter with FWHMs set as $5.5$~mm and $7.0$~mm for transaxial and axial components, respectively -- close to the PSFs estimated in the dedicated work \cite{Moskal2021a};
\item[--] "original": our algorithm, with a realistic analytical RM.
\end{itemize}

\subsection{NEMA IEC phantom}
\begin{figure}[!t]
\centering
\includegraphics[width=22pc]{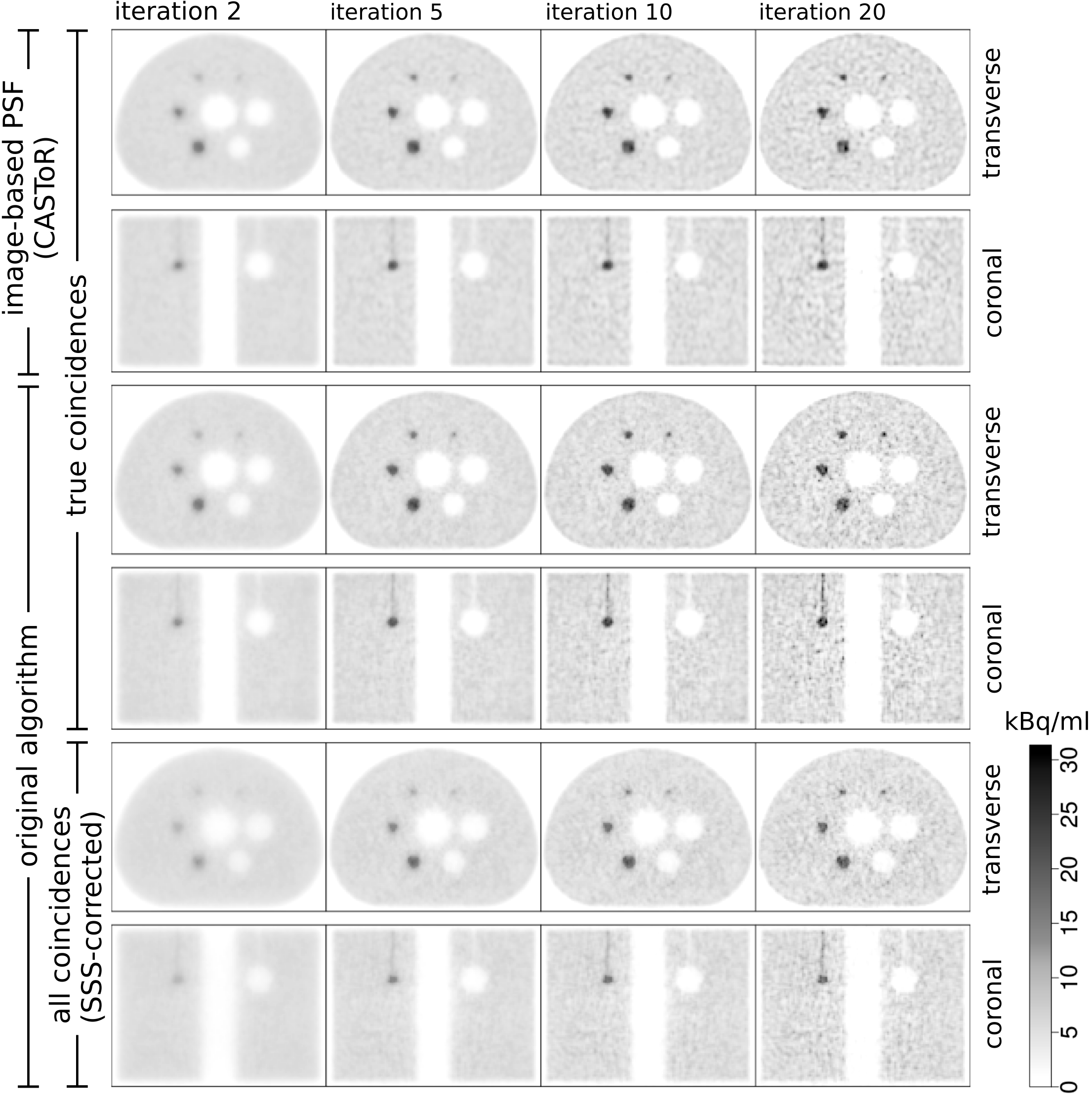}
\caption{Selected cross-sections of the reconstructed NEMA IEC phantom for the simulated $35$-s scan, using various TOF MLEM implementations and coincidence types present in the datasets.}
\label{Fig:IEC_recons_35s}
\end{figure}

\begin{figure}[!t]
\centering
\includegraphics[width=22pc]{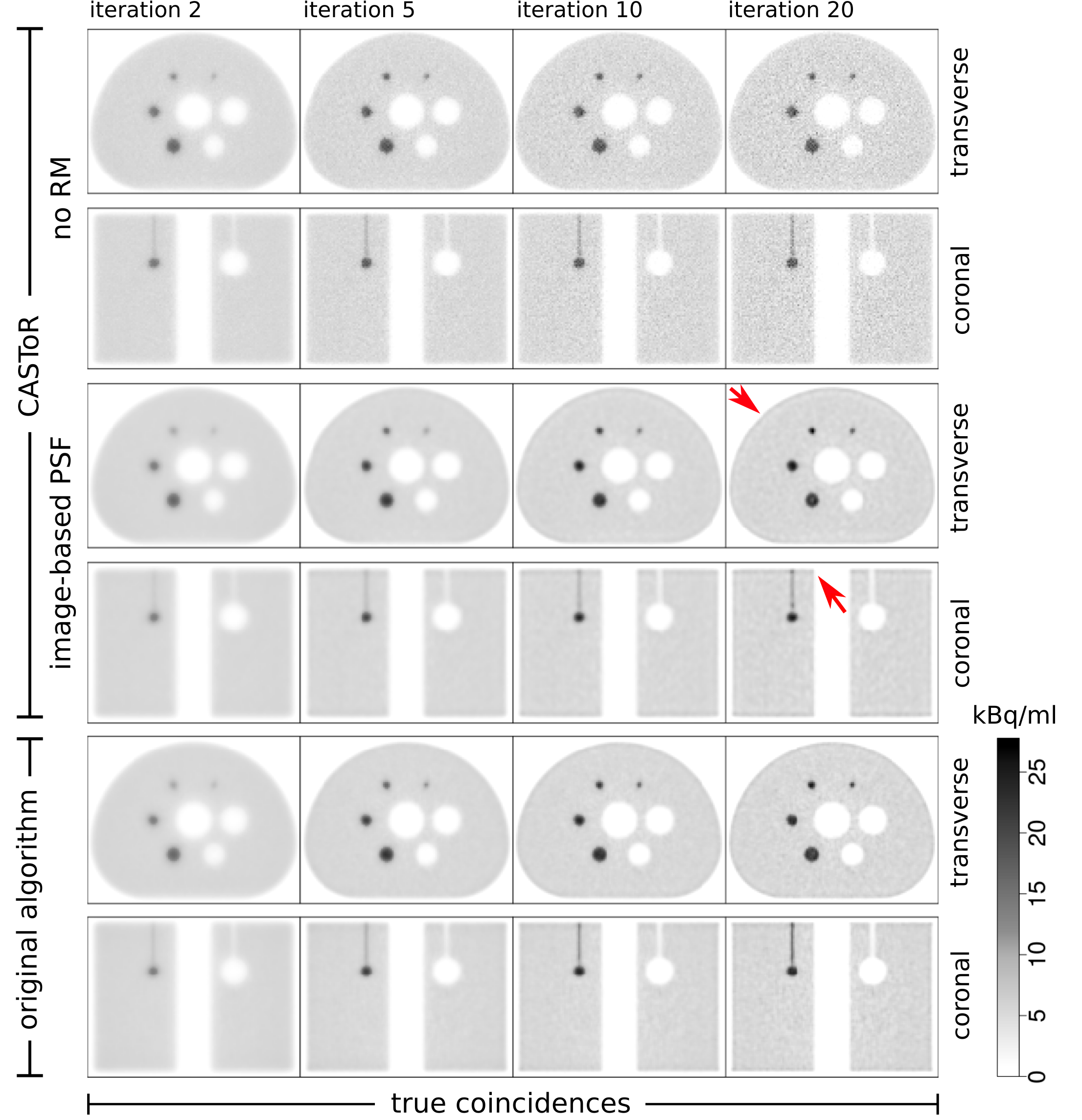}
\caption{Selected cross-sections of the NEMA IEC phantom, reconstructed for the simulated $500$-s scan (true events only) using various TOF MLEM implementations. The arrows point where Gibbs overshoot is visible.}
\label{Fig:IEC_recons_500s}
\end{figure}

Fig.~\ref{Fig:IEC_recons_35s} and Fig.~\ref{Fig:IEC_recons_500s} show the transverse ($z=37$~mm) and coronal ($y=0$~mm) cross-sections of the reconstructed NEMA IEC phantom, obtained using different RMs for $35$-s and $500$-s scans, respectively. The original TOF MLEM produces slightly better local contrast than the PSF from CASToR, more visible in the first figure, indicating superior resolution recovery. The lack of RM substantially increases noise (top row in Fig.~\ref{Fig:IEC_recons_500s}), while shift-invariant PSF produces more notable artefacts at the edges (Gibbs overshoot) enhanced at higher iterations for the larger dataset, in particular along the axial direction, not observed for our implementation. 

\begin{figure}[!t]
\centering
\includegraphics[width=25pc]{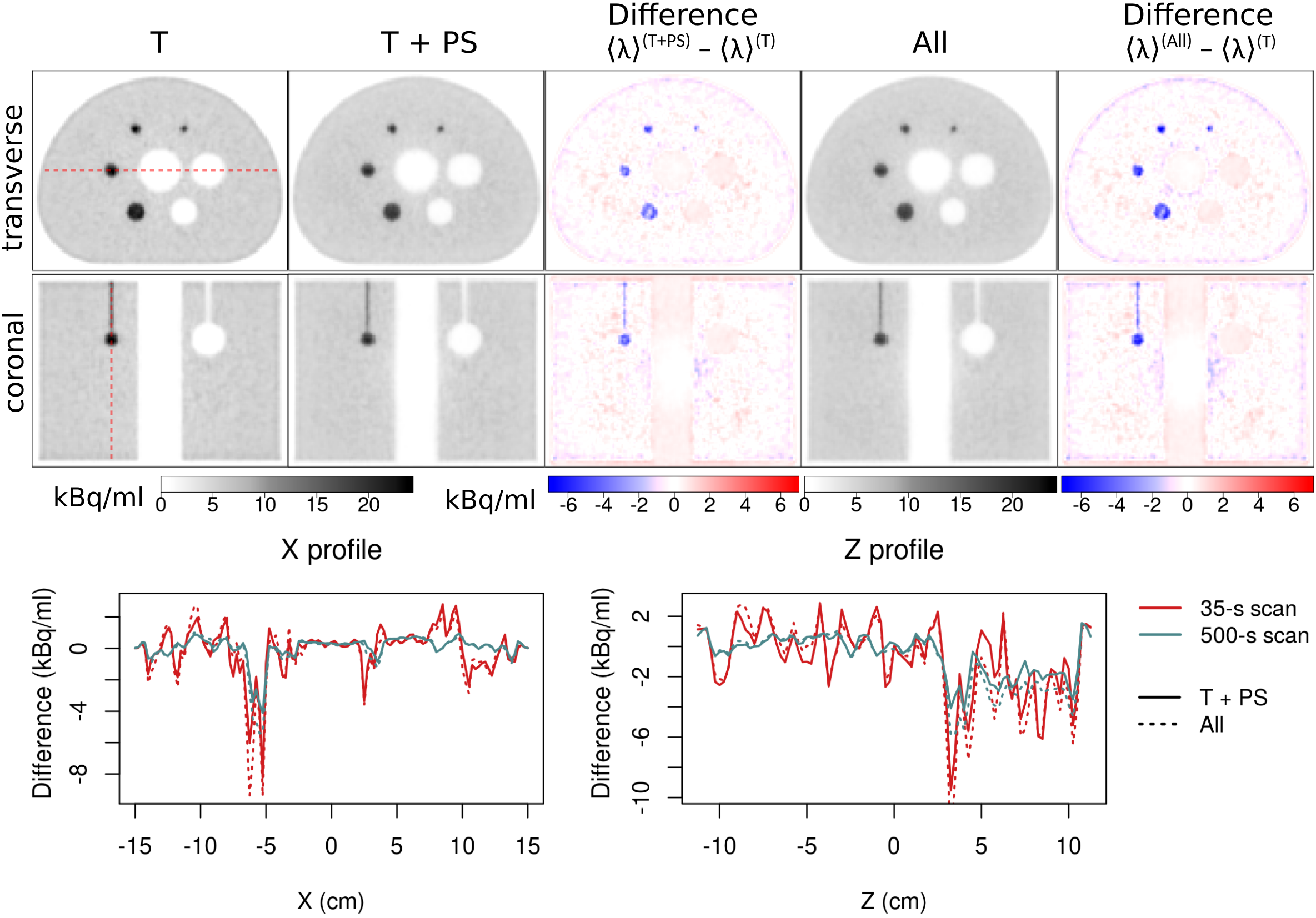}
\caption{Top: NEMA IEC activity concentrations $\langle\bm\lambda\rangle$ and the difference images, reconstructed for the true (T), true + phantom scattered (T + PS) or all types (All) of coincidences ($15$-th iteration, $500$-s scan). Bottom: 1D difference profiles built along the dashed lines shown in cross-sections.}
\label{Fig:Diffs_IEC_15it}
\end{figure}

Fig.~\ref{Fig:IEC_recons_35s} (bottom row) also presents the results for a dataset with all additive contributions ("all coincidences") and scatter correction applied. Very similar images were obtained for the data without random events and scattered in the detector (see Table~\ref{Tab:CTypes}), yet also inferior to the case of only true coincidences -- with lower activity concentration in the hot spheres. To distinguish the outcomes, we checked the difference images between three cases: true (T), true + phantom scattered (T + PS) and all types (All) of coincidences taken for the reconstruction. As shown in Fig.~\ref{Fig:Diffs_IEC_15it}, the additive factors mostly affect the hot regions (see 1D profiles). A minor axial inconsistency is visible at the centre, in particular for the difference between the reconstructed activities $\langle\bm\lambda\rangle^\text{(All)}-\langle\bm\lambda\rangle^\text{(T)}$ in coronal slice. We explored it is unrelated to the geometrical pre-filtering and cannot be eliminated by re-normalisation of $\hat{s}_{i_\epsilon}$. Presumably, a further refinement of the SSS method can help -- either energy window adjustment (we used the default $350$~keV -- $650$~keV), higher sinogram resolution and/or TOF information included.

\subsection{Static XCAT phantom}
\begin{figure*}[!t]
\centering
\includegraphics[width=402pt]{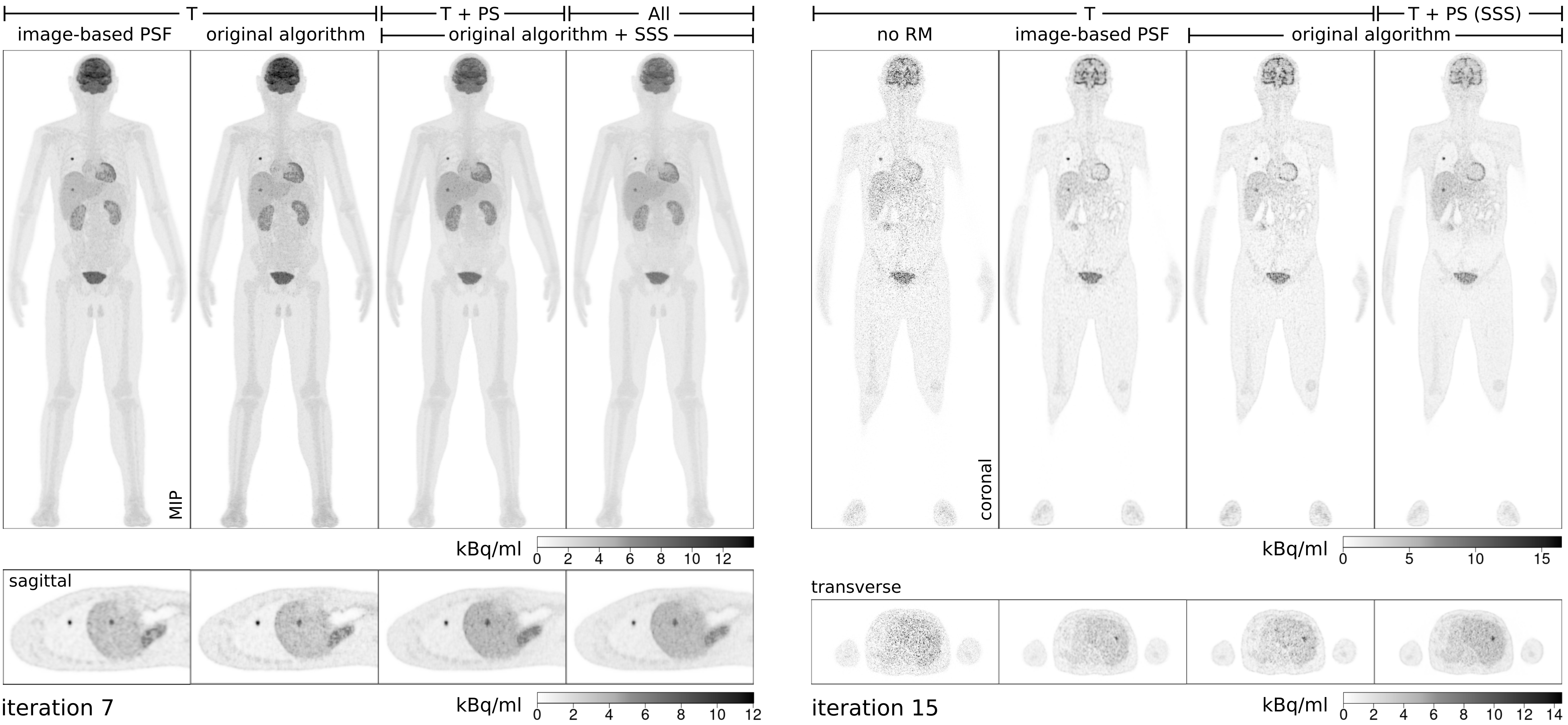}
\caption{Selected MIPs and cross-sections of the XCAT, reconstructed by various TOF MLEM implementations and coincidence types included in the dataset. The slices were selected to cover the $12$-mm lesions: $x=90$~mm (sagittal, torso part), $y=11.25$~mm (coronal) and $z=371.25$~mm (transverse, across the liver). T -- true, T + PS -- true + phantom scattered, All -- all types of coincidences.}
\label{Fig:XCAT_recs}
\end{figure*}

Fig.~\ref{Fig:XCAT_recs} presents the maximum intensity projections (MIPs) along the $Y$-axis and the various cross-sections of the simulated XCAT phantom, reconstructed after the $7$-th and $15$-th iterations, depending on the TOF MLEM setup and types of coincidences that constituted the dataset. Similarly to the NEMA IEC, the difference between the image-based PSF (CASToR) and our method is subtle for true events, mostly seen around lesions, brain area (coronal slice) and far from the centre of AFOV (palms, legs).

\begin{figure}[!t]
\centering
\includegraphics[width=25pc]{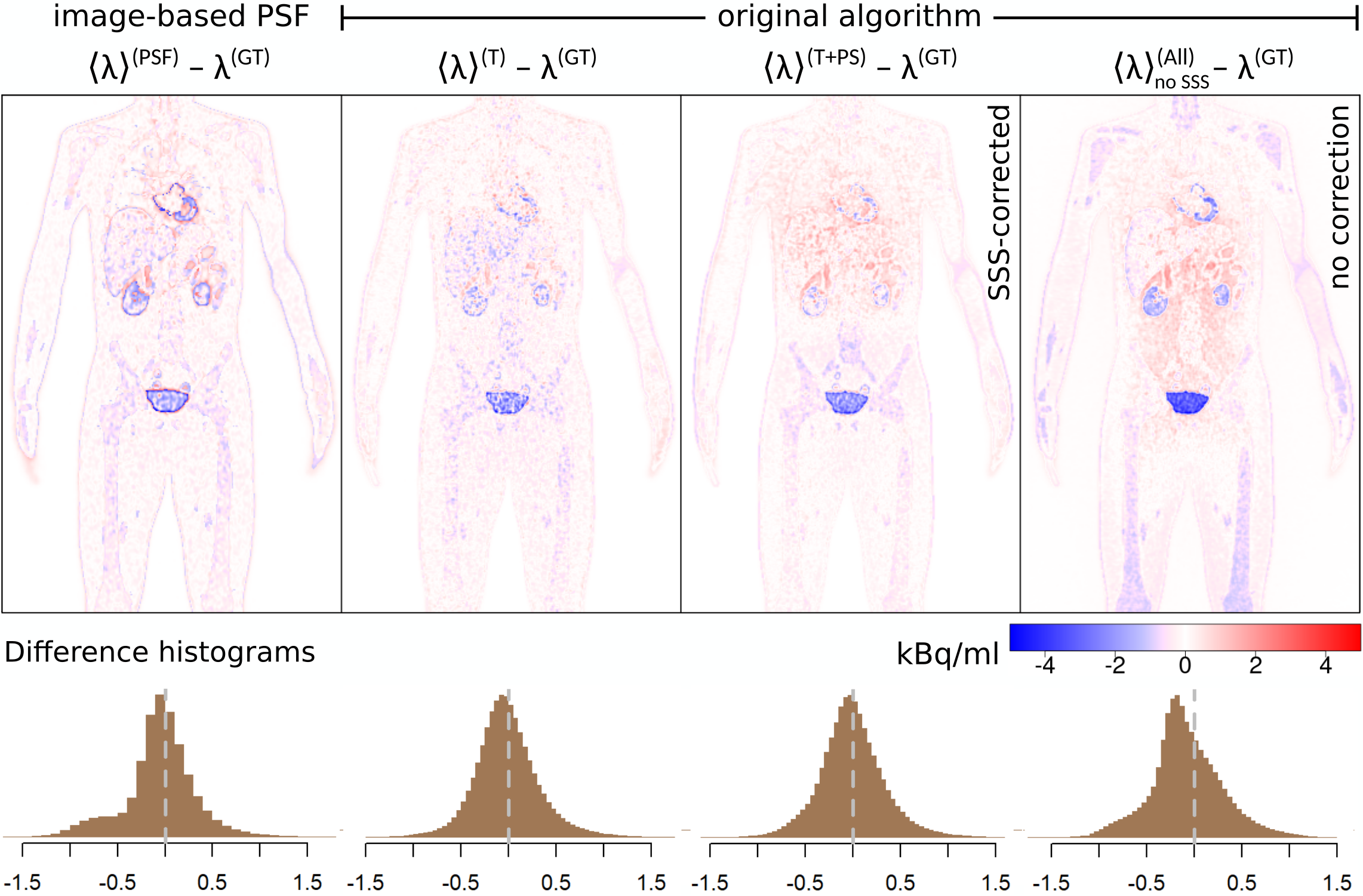}
\caption{Partial coronal cross-sections ($y=-6.25$~mm, covering arms) of the $\langle\bm\lambda\rangle-\bm\lambda^\text{(GT)}$ differences for various reconstructed XCAT activity concentrations after the $7$-th iteration (top) and the corresponding histograms built for the voxels under the condition $\bm\lambda^\text{(GT)}>0$ (bottom).}
\label{Fig:XCAT_Errs}
\end{figure}

More details can be extracted from the differences between the reconstructed activities $\langle\bm\lambda\rangle$ and GT -- $\bm\lambda^\text{(GT)}$. The examination of Fig.~\ref{Fig:XCAT_Errs}, top around the torso region indicates that the PSF correction exhibits worse edge preservation, even compared with the data containing additive components. The realistic SRM modelling leads to higher spatial resolution and more accurate reconstruction of the areas far from the centre of the scanner. The histograms at the bottom of Fig.~\ref{Fig:XCAT_Errs} are as well more symmetric for our method. In the presence of phantom scattered events (T + PS), there is some noise left after SSS-correction, but far smaller than in the uncorrected image (on the right) and with a rectified bias in the histogram. It is important to note that the differences $\langle\bm\lambda\rangle^\text{(All)}-\bm\lambda^\text{(GT)}$ and the histogram for the scatter-corrected data are close to the ones for T + PS, despite the presence of randoms (not shown).

\subsection{Image quality}
\begin{figure}[!t]
\centering
\includegraphics[width=24pc]{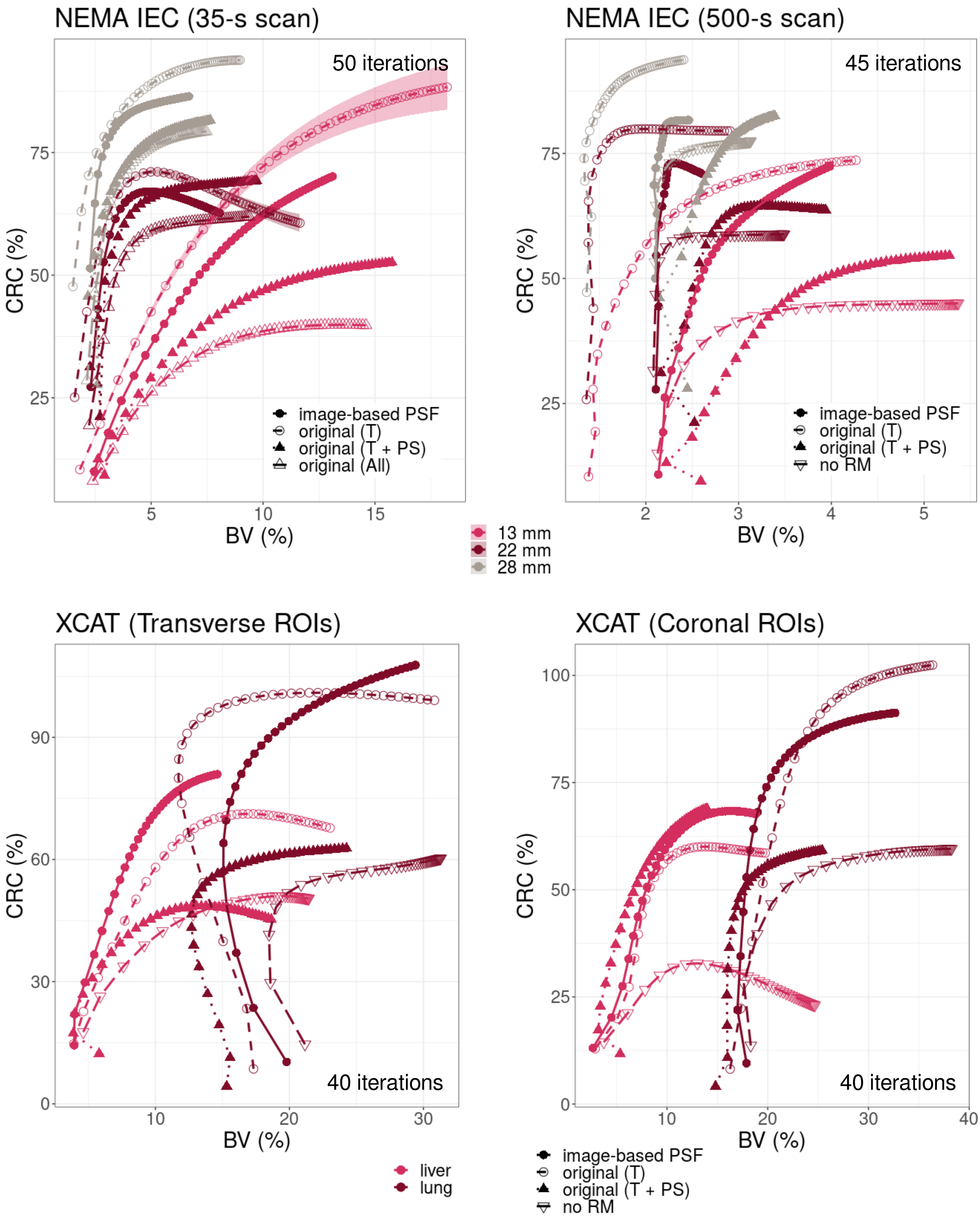}
\caption{Image quality parameters of the NEMA IEC (top) and XCAT (bottom), reconstructed by the TOF MLEM, depending on the RM, coincidence types in the dataset and ROI definition methods (XCAT). Ribbons denote the SDs of the CRC that reflect various ROI choice for the NEMA IEC (top left).}
\label{Fig:IQs}
\end{figure}

Fig.~\ref{Fig:IQs}, top presents the estimated $\text{CRC}(\text{BV})$ dependencies for three NEMA IEC spheres (two hot and one cold), calculated for $35$-s (left) and $500$-s (right) scans, utilising diverse TOF MLEM variants. Considering the $\text{CRC}$ variance of the ROI choice (shown for the "original (T)" as ribbons), the curves for the realistic RM (\ref{Eq:SRManalyticalNew}) are superior and closer to the ideal point $\text{CRC}=1, \text{BV}=0$, compared with a single Siddon projector used by CASToR (only true events), with or without a stationary Gaussian PSF kernel. The original algorithm also converges faster, but reaches higher $\text{BV}$ for the NEMA IEC than the reference (with PSF), especially for the shorter scan, which points at a need for regularisation. We also observe a $\text{CRC}$ degradation for the $22$-mm sphere (NEMA IEC) and the $12$-mm lesion in the liver (XCAT), not directly related to the TOF MLEM setup and PET data.

Scatter-corrected data exhibit worse local contrast, as expected, further suppressed in the presence of random and detector scattered events (Fig.~\ref{Fig:IQs}, top left). We see an inconsistency, though, for the two methods of ROI allocation inside the liver XCAT region (brighter solid triangles, at the bottom). Presumably, an alternate approach should be utilised for the $\text{CRC}(\text{BV})$ assessment, different from the one proposed in \cite{Zhang2017}. 

Although we can distinguish the outcomes for various TOF MLEM setups, image quality assessment may be further refined in terms of ROI choice and/or averaging over multiple measurements/simulations, as required by NEMA.

\subsection{Ground truth metrics}
\begin{figure}[!t]
\centering
\includegraphics[width=25pc]{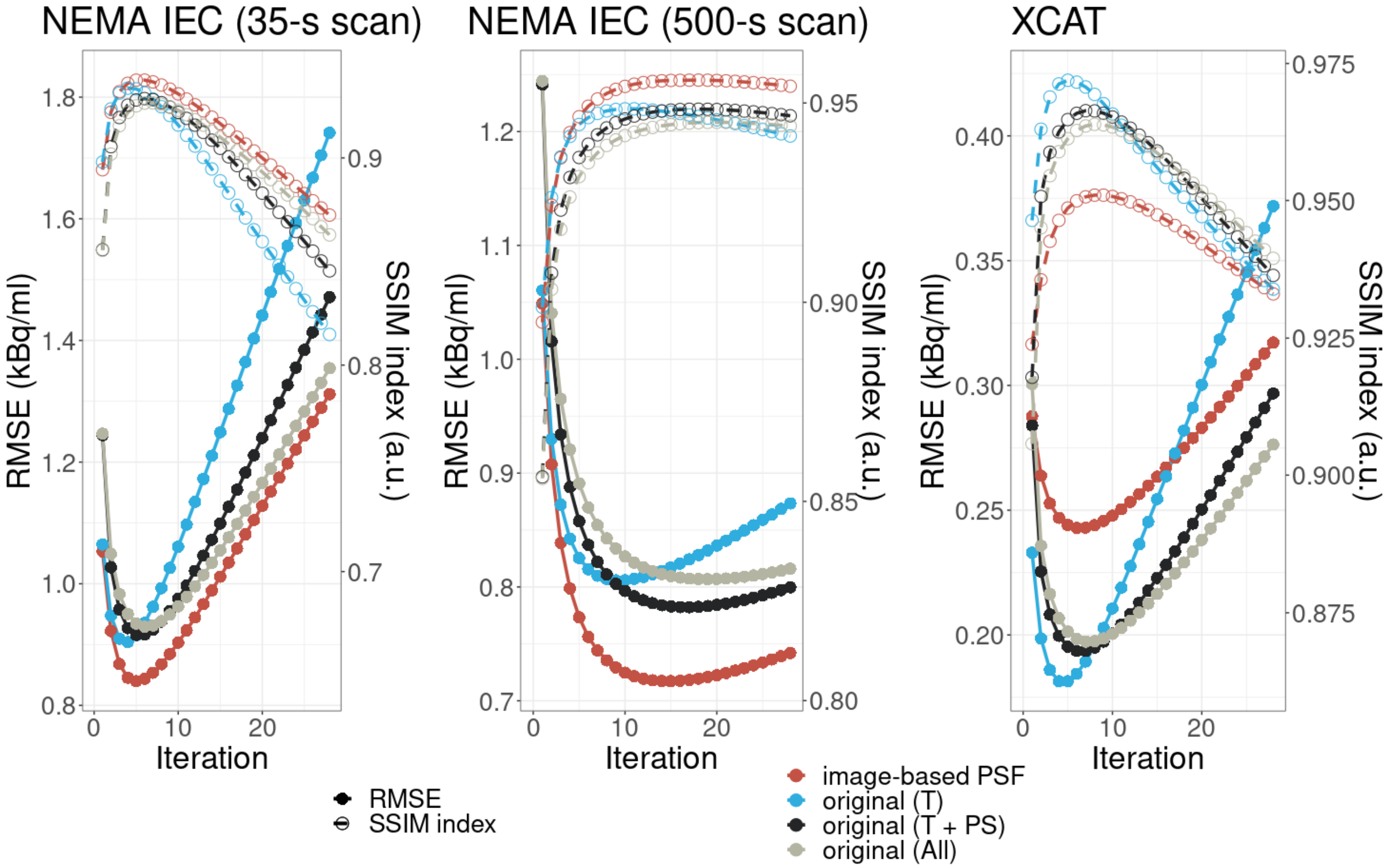}
\caption{Evolution of $\text{RMSE}_\text{GT}$ and $\text{SSIM}_\text{GT}$ over 30 iterations, depending on the phantom, RM, time of scan and coincidence types in the dataset.}
\label{Fig:Metrics}
\end{figure}

The results for $\text{RMSE}_\text{GT}$ and $\text{SSIM}_\text{GT}$, estimated for different RMs and datasets are collected in Fig.~\ref{Fig:Metrics}. The absence of RM leads to much worse GT metrics (not depicted). An interesting observation is that the shift-invariant PSF, applied to the NEMA IEC, results in better $\text{RMSE}_\text{GT}$ and $\text{SSIM}_\text{GT}$, compared to our model (true events), especially for the $500$-s scan. Another unexpected outcome is that the NEMA IEC, reconstructed from the scatter-corrected data comprising incomplete additive factors (T + PS), reaches lower $\text{RMSE}_\text{GT}$ than the "true only" case (not seen for $\text{SSIM}_\text{GT}$). Both issues appear to be phantom-related, since the opposite -- more logical -- regularities are observed for the XCAT: the best metrics calculated for the original RM model and the worse -- for a stationary PSF convolution from CASToR.

The best (minimal) $\text{RMSE}_\text{GT}$ is generally shifted to lower iterations compared to the best (maximal) $\text{SSIM}_\text{GT}$. The curves are more consistent for the latter (see Fig.~\ref{Fig:Metrics}), proving that the SSIM index outperforms RMSE in measuring the quality of natural images across a wide variety of distortions \cite{Wang2004}.

\section{Discussion}
The results reported so far for the novel TOF MLEM are overall promising, but more tests are required to clarify the difference between our implementation and the reference. We shall therefore discuss the phantom impact and the kernel role -- a convolutional PSF or 1D functions $H_\text{TOF}(\cdot)$ and $H_Z(\cdot)$.

\subsection{Phantom role for the $\text{RMSE}_\text{GT}$ and $\text{SSIM}_\text{GT}$ assessment}
As seen in Fig.~\ref{Fig:Metrics}, the shift-invariant PSF correction produced the best $\text{RMSE}_\text{GT}$ and $\text{SSIM}_\text{GT}$ for a longer NEMA IEC scan. This result may mislead unless we take into account the specifics of the phantom -- a relatively small and located at the centre of the AFOV. A fixed PSF does not enhance smearing, and the majority of the NEMA IEC volume has uniform activity, $4\times$ lower than in the hot spheres, contributing the most to the metrics. In other words, the smoother the "background volume" is, the better numbers we get. In comparison, the activity distribution in XCAT is more "diverse", hence the curves are more separable from each other in Fig.~\ref{Fig:Metrics}.

The quasi-asymptotic convergence seen in Fig.~\ref{Fig:Metrics} may as well indicate that the image-based PSF acts like a regularisation. If we as well apply image-domain penalisation techniques, the outcomes for the novel TOF MLEM would be much better, as already studied in a dedicated work \cite{ShopaRAP2021}. 

\subsection{Kernel shape and size}
Given the relatively low count of detected emissions or the presence of additive factors, one can simplify the RM kernels as a trade-off between the performance, noise and image quality. To start with, it is rewarding to redefine the polynomial fitting functions $\chi_i\left(l^\parallel, l^\perp,\theta\right)$ (Fig.~\ref{Fig:SRMinDJPET}, b) in log-scale and exclude some irrelevant terms (see Appendix).

\begin{figure}[!t]
\centering
\includegraphics[width=22.5pc]{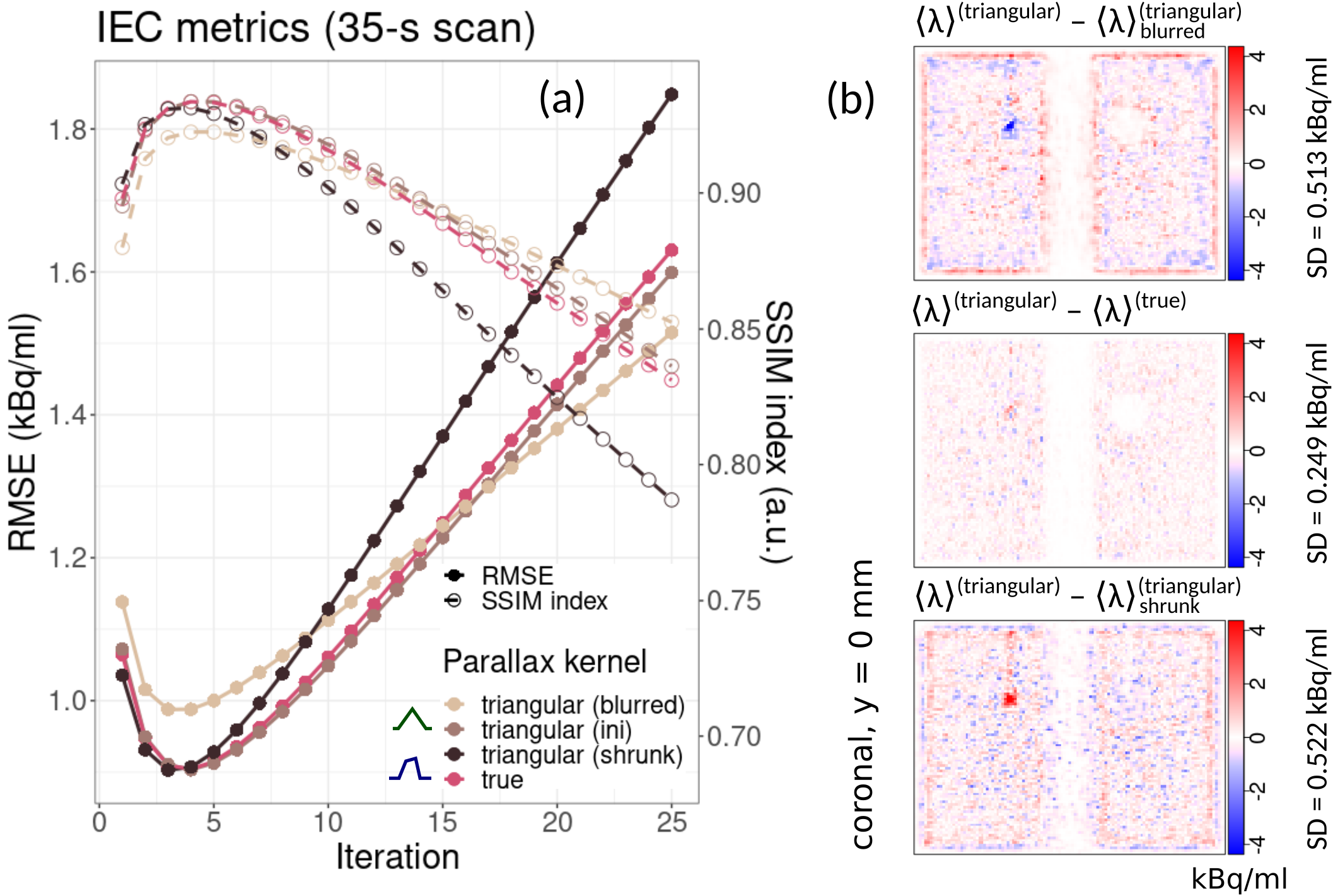}
\caption{Evolution of the GT metrics over iterations (a) and the differences between the reconstructed NEMA IEC scan ($13$-th iteration) (b), depending on the RM kernel parameters.}
\label{Fig:KernImpact}
\end{figure}

Using the $35$-s NEMA IEC dataset, we as well explored the role of the TOF and $Z$-kernels, imposing a simplified model for the convolution $H^\text{CRT}_Z(\cdot)*H^\text{prlx}_Z(\cdot)$. Fig.~\ref{Fig:KernImpact}, a shows the iterative progress of $\text{RMSE}_\text{GT}$ and $\text{SSIM}_\text{GT}$, estimated for the precisely calculated $H^\text{prlx}_Z(\cdot)$ ("true") and using a simplified (triangular) kernel with the same FWHM. Although the difference is barely seen, a $2\times$ boost in performance was achieved. This indicates a minor role of the kernel shape, confirmed by the $\text{CRC}(\text{BV})$ dependencies, also close to each other (not shown).

Two additional curves are depicted for the triangular model in Fig.~\ref{Fig:KernImpact}, a  -- with the SDs $\sigma_\text{TOF}$ and $\sigma_Z$ deliberately enlarged ("blurred") or reduced ("shrunk") by $50\%$. It appears that such adjustments deteriorate the results more significantly than changing the kernel shape, which is seen in Fig.~\ref{Fig:KernImpact}, b -- for the differences between the images at the $13$-th iteration, when the GT metrics are close to each other. That also underlines the need for the accurate assessment of the J-PET resolution, which the optimal $\sigma_\text{TOF}$ and $\sigma_Z$ directly depend upon.

\subsection{PSF adjustment in CASToR}
Since the activity concentrations reconstructed by the original TOF MLEM are close to the results for the CASToR with a convolver, we have tested a "de-tune" of the image-based PSF to check how it affects the outcome. It was done by a $\times1.5$ reduction or enhancement of FWHMs used for the Gaussian convolutional kernel. The corresponding values covered the span $3.3$~mm -- $7.5$~mm in the transverse and $5.0$~mm -- $11.25$~mm -- in the axial directions. 
\begin{figure}[!t]
\centering
\includegraphics[width=23.5pc]{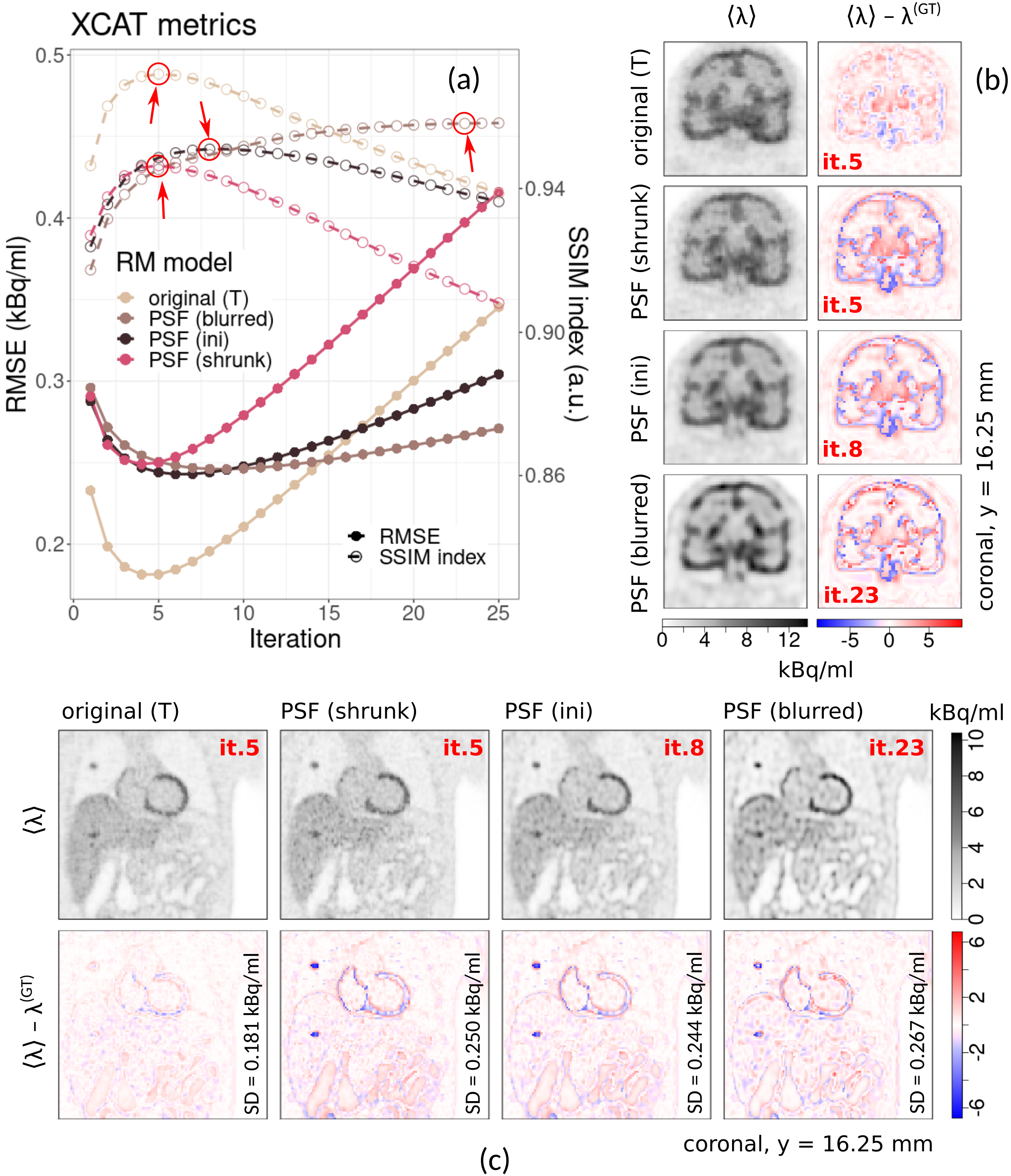}
\caption{Evolution of the GT metrics over $25$ iterations (a), the exemplary coronal cross-sections around the brain (b) and torso (c) of the reconstructed XCAT (iterations for the highest $\text{SSIM}_\text{GT}$ pointed by red arrows), along with the relevant difference images $\langle\bm\lambda\rangle-\bm\lambda^\text{(GT)}$, depending on the RM model.}
\label{Fig:PSFImpact}
\end{figure}

Selected results are aggregated in Fig.~\ref{Fig:PSFImpact}. The close-up brain and torso areas of the coronal XCAT cross-sections are compared for the iterations with the highest $\text{SSIM}_\text{GT}$. SDs for $\langle\bm\lambda\rangle-\bm\lambda^\text{(GT)}$  and $\text{RMSE}_\text{GT}$ reach the absolute minimum for the "default" FWHMs, which corresponds to the optimal PSFs estimated in our earlier study \cite{Moskal2021a}. The reduced ("shrunk") kernel exhibit higher contrast yet adds more noise than our algorithm produces. With further increase of the FWHMs, PSF correction resembles regularisation: $\text{SSIM}_\text{GT}$ asymptotically converges after $20$-th iteration. However, that also worsens the contrast and has the bad edge preservation, visible in the difference images (Fig.~\ref{Fig:PSFImpact}, b-c). Similar regularities were observed for the images reconstructed for the same iteration. 

The analysis proves that the initial kernel choice was fair and our RM model is superior to the fixed image-based PSF convolution. It could be beneficial for additional regularisation, preventing unexpected artefacts to emerge.

\subsection{Algorithm limitations and potential upgrade}
The main motivation for using log-polynomial $\chi_i\left(l^\parallel_j, l^\perp_j, \theta \right)$ is the axial continuity of J-PET scintillators, which would require a major redesign for a multi-ring scanner (under testing in GATE). Besides, a voxel-by-voxel SRM recalculation across a volume around an annihilation point is obviously slower than the ray-driven Siddon projector used in CASToR, which is a challenge to be resolved the future. As already shown, a rigorous RM might be simplified to boost performance, but its optimisation may require a dedicated study (see \cite{Shopa2021}). 

On the other hand, we predict a more controllable behaviour of the realistic SRM for further expansion -- to cover all additive corrections, positron range and non-collinearity, or with regularisation applied (some tests reported in \cite{ShopaRAP2021}). The detector blur $\textbf{M}_\text{det.res}$ is modelled fairly accurately, proved by image quality and edge preservation (Fig.~\ref{Fig:XCAT_Errs}) -- superior to the image-based PSF even in the presence of the errors $\hat{\textbf{b}}$. Also, as Table~\ref{Tab:CTypes} indicates, one cannot neglect the scattering of $\gamma$-photons inside the detection modules, including WLS-strips (see Figs.~\ref{Fig:DJPET}). This substantiates the need for a more advanced scatter correction method, other than a phantom-based SSS. 

It is theoretically possible to embed the analytical RM in CASToR, which allows using an external SRM -- a promising perspective, providing it has no tools so far to directly account for attenuation and scattering in passive elements like WLS.

Finally, we foresee the TOF MLEM application in multi-photon imaging \cite{Moskal2021}. Due to a much larger bin space, it is unpractical to employ a Monte Carlo based SRM modelling, but multiple projectors could be used to account for $\textbf{M}_\text{det.res}$ in a similar way on-the-fly. Another problem concerns the integral (\ref{Eq:SensAttSum}), which becomes too expensive to calculate in terms of attenuation factors. This appears to be a major challenge to be resolved in the future.

\section{Conclusions}
We performed a dedicated research of the original TOF MLEM tomographic reconstruction algorithm, designed for the next generation TB J-PET scanner, using the simulated scans of the NEMA IEC and static XCAT phantoms under conditions, consistent with a similar studies and/or scanner geometries similar to the currently used clinical PET systems.

Having capitalised on the J-PET design, used here as an example of the multi-layer system, we modelled the shift-variant part of SRM as a result of 2D Monte Carlo simulations and log-polynomial fitting applied to each pair of scintillator strips. Further extension of SRM to account for TOF information, axial smearing and parallax correction allowed to build a realistic RM for the scanner, as well as utilise it for the calculation of sensitivity and attenuation factors. The shift-variant SRM can also be estimated on-the-fly by multiple projectors, potentially applicable to multi-photon imaging. 

The novel algorithm was tested on the simulated data of various acquisition times, post-filtered to include only true, true and phantom scattered, or all types of coincidences. The data, reconstructed by TOF MLEM, was compared to the reference method from the CASToR software with no or a generic image-based RM. The results showed the reliability of the proposed RM, visually and quantitatively -- from the analysis of image quality, GT metrics and difference images. The presence of the additive factors produced the outcomes, slightly inferior to the case of only true events, yet outperforming the reference in terms of edge preservation. 

The benchmark analysis of the scripts written in {\small\textsf{R/Rcpp}} was not conducted. However, the software is adapted to run in parallel and can potentially be migrated to dedicated platforms such as graphics processing units. The performance can be further boosted by using simpler kernel components for RM. Along with exploring alternative functions to fit the shift-invariant SRM part, this could become a rewarding strategy for further RM upgrade to account for non-collinearity and positron range, as well as to include penalisation or more advanced additive corrections.

\section*{Acknowledgements}
This work was supported by Foundation for Polish Science through TEAM POIR.04.04.00-00-4204/17, the
National Science Centre, Poland (the grants No. 2021/42/A/ST2/00423, 2021/43/B/ST2/02150 and 2020/37/N/NZ7/04106) and by the Ministry of Education and Science under the grant No. SPUB/SP/530054/2022. It has also been supported by a grant from the SciMat and qLife Priority Research Areas under the Strategic Programme Excellence Initiative and by the Jagiellonian University (project CRP/0641.221.2020).


\appendix
\section*{Appendix: SRM fitting function}
To represent the shift-variant SRM elements $\chi_{i j}$, we model their components in detector space as analytical functions $\chi_i\left(l^\parallel_j, l^\perp_j, \theta \right)$ of the obliqueness $\theta$ and transverse variables $l^\parallel_j$ and $l^\perp_j$, denoting the $j$-th voxel position along a LOR and in a perpendicular direction, respectively, (see (\ref{Eq:SRManalyticalNew}) and Fig.~\ref{Fig:SRMinDJPET}). We explore that the Monte Carlo simulated data is best fitted by the {\small\textsf{R}} regression function {\small\textsf{lm()}} using a log-polynomial expression with $35$ free coefficients $A_0\ldots A_{34}$:
\begin{equation}
\label{Eq:A_SRMformula}
\begin{aligned}
&\log\left[\chi_i\left(l^\parallel_j, l^\perp_j, \theta \right)+1\right] = \exp[ \\
&A_0 + A_1 l^\parallel_j + A_2 \vert l^\parallel_j \vert + A_3 (l^\parallel_j)^2 + \\
&A_4 l^\perp_j + A_5 (l^\perp_j)^2 + A_6 (l^\perp_j)^3 + A_7 (l^\perp_j)^4 + A_8 (l^\perp_j)^5 + \\
&A_9\tan{\theta} + A_{10}\tan^2{\theta} + \\
&A_{11} l^\parallel_j l^\perp_j + A_{12} l^\parallel_j (l^\perp_j)^2 + A_{13} l^\parallel_j (l^\perp_j)^3 
	+ A_{14} l^\parallel_j (l^\perp_j)^4 + \\
&A_{15} \vert l^\parallel_j \vert l^\perp_j + A_{16} \vert l^\parallel_j \vert (l^\perp_j)^2 
	+ A_{17} \vert l^\parallel_j \vert (l^\perp_j)^3 + A_{18} \vert l^\parallel_j \vert (l^\perp_j)^4 + \\
&A_{19} (l^\parallel_j)^2 l^\perp_j + A_{20} (l^\parallel_j)^2 (l^\perp_j)^2 + A_{21} (l^\parallel_j)^2 (l^\perp_j)^3 + \\
&A_{22}l^\parallel_j l^\perp_j \tan{\theta} + A_{23}l^\parallel_j (l^\perp_j)^2 \tan{\theta} 
	+ A_{24}l^\parallel_j (l^\perp_j)^3 \tan{\theta} + \\
&A_{25}\vert l^\parallel_j \vert l^\perp_j \tan{\theta} 
	+ A_{26}\vert l^\parallel_j \vert (l^\perp_j)^2 \tan{\theta}
	+ A_{27}\vert l^\parallel_j \vert (l^\perp_j)^3 \tan{\theta} + \\
&A_{28}(l^\parallel_j)^2 l^\perp_j \tan{\theta} + A_{29}(l^\parallel_j)^2 (l^\perp_j)^2 \tan{\theta} + \\
&A_{30} l^\parallel_j l^\perp_j \tan^2{\theta} 
	+ A_{31} l^\parallel_j (l^\perp_j)^2 \tan^2{\theta} + \\
&A_{32}\vert l^\parallel_j \vert l^\perp_j \tan^2{\theta} 
	+ A_{33}\vert l^\parallel_j \vert (l^\perp_j)^2 \tan^2{\theta} + \\
&A_{34}(l^\parallel_j)^2 l^\perp_j \tan^2{\theta} 
].
\end{aligned}
\end{equation}

The point $(l^\parallel_j=0, l^\perp_j=0)$ reflects the LOR centre. The use of logarithmic scale and a module $\vert l^\parallel_j \vert$ substantially reduced the RMSE, while some terms of the polynomial (containing $(l^\parallel_j)^3, \tan^3{\theta}$ or higher) have been excluded due to a high $p$-value (available from the fitting summary in {\small\textsf{R}} \cite{RCoreTeam2020}).

\end{document}